\documentclass[fleqn,usenatbib]{mnras}

\usepackage{newtxtext,newtxmath}
\usepackage{xcolor}

\usepackage[T1]{fontenc}

\DeclareRobustCommand{\VAN}[3]{#2}
\let\VANthebibliography\thebibliography
\def\thebibliography{\DeclareRobustCommand{\VAN}[3]{##3}\VANthebibliography}

\usepackage{graphicx}	
\usepackage{amsmath}	
\usepackage{soul}

\title[Wind and jet from accretion flows]{The properties of wind and jet from a super-Eddington  accretion flow around a supermassive black hole}

\author[H. Yang, et al. ]{
Hai Yang$^{1,2}$\thanks{E-mail: hyang@shao.ac.cn (HY); fyuan@shao.ac.cn (FY); lixindai@hku.hk (LD)},
Feng Yuan$^{1,2}$,
Tom Kwan$^{3}$,
and Lixin Dai$^{3}$
\\
$^{1}$Shanghai Astronomical Observatory, Chinese Academy of Sciences; 80 Nandan Road, Shanghai 200030, China\\
$^{2}$University of Chinese Academy of Sciences; 19A Yuquan Road, Beijing 100049, China\\
$^{3}$Department of Physics, University of Hong Kong, Pokfulam Road, Hong Kong, China
}

\date{Accepted 2023 May 09. Received 2023 April 12; in original form 2022 November 22}

\pubyear{2023}

\begin{document}
\label{firstpage}
\pagerange{\pageref{firstpage}--\pageref{lastpage}}
\maketitle

\begin{abstract}
Wind and jet are important medium of AGN feedback thus it is crucial to obtain their properties for the feedback study. In this paper we investigate the properties of wind and jet launched from a magnetized super-Eddington accretion flow around a supermassive black hole. For this aim, we have performed  radiation magnetohydrodynamical simulation of a magnetically arrested super-Eddington accretion flows. We then have analyzed the simulation data by the ``virtual particle trajectory'' approach and obtained the mass flux, poloidal and toroidal velocities, and mass-flux-weighted momentum and energy fluxes of wind and jet. The mass flux is found to be 2-6 times higher than that obtained based on the time-averaged streamline method widely used in literature. The momentum flux of wind is found to be larger than that of jet, while the total energy flux of jet is at most 3 times larger than  that of wind. These results are similar to the case of hot accretion flows and imply that winds likely play a more important role than jet in AGN feedback. The acceleration mechanism of wind and jet is analyzed and found to be dominated by Lorentz force rather than radiation force.  
\end{abstract}



\section{Introduction} \label{sec:intro}

There are strong observational evidences for the co-evolution between the supermassive black hole and its host galaxy and AGN feedback is believed to be the responsible mechanism \citep{Fabian2012,Kormendy2013Ho,ostriker2017}. The feedback is mediated by three kinds of ``outputs'' from the AGN, namely radiation, wind, and jet. In the present paper we focus on  wind and jet\footnote{Note that in the literature, many papers do not clearly discriminate jet and wind and these two names are often ``mix-use''. In the black hole accretion theory, wind and jet are intrinsically different. The jet is closer to the spin axis of the black hole. The velocity of jet is much faster than wind while its opening angle is much smaller than wind. See Section \ref{definition} for our detailed definitions of wind and jet.}. There have been many works studying the roles of jet and wind in AGN feedback  at various scales, from galaxy \citep{2009ApJ...699...89C,2010ApJ...722..642O,2010ApJ...717..708C,2015ARA&A..53..115K,2018ApJ...857..121Y,2018ApJ...864....6Y,2019ApJ...872..167G,2021A&A...654A...8N, Chowdhury22}, galaxy cluster \citep{2014ApJ...789...54L,2016ApJ...829...90Y,2017MNRAS.470.4530W,2018MNRAS.473.1332G}, and cosmological scales \citep{2018MNRAS.479.4056W,2019MNRAS.486.2827D,2014MNRAS.444.1453D}.  For example, \citet{2018ApJ...857..121Y} included feedback by wind and radiation from AGNs and found that wind plays a dominant role in controlling
the star formation and black hole growth, although radiative
feedback cannot be neglected. Obviously, a crucial information for this study is the properties of jet and wind, including their mass flux, velocity, and fluxes of momentum and energy, as a function of AGN accretion rate and black hole mass. 

Depending on the value of the temperature of the accretion flow, black hole accretion is divided into two modes, i.e., cold and hot ones, corresponding to two feedback modes. These two modes are determined by the mass accretion rate of the black hole. When $\dot{M}\la 2\% \dot{M}_{\rm Edd}$ ($\dot{M}_{\rm Edd}=10 L_{\rm Edd}/c^2$ is defined as the Eddington accretion rate, with $L_{\rm Edd}$ being the Eddington luminosity), the accretion is in the hot mode \citep{Yuan2014Narayan}. The study of wind in a hot accretion flow can be traced back to \citet{1999MNRAS.310.1002S}, which is the first global numerical simulation of black hole accretion. This work finds that the mass accretion rate is not a constant of radius, but decreases inward. \citet{1999MNRAS.303L...1B} speculated that the inward decrease of accretion rate is because of mass loss via wind \citep[see also][]{1994ApJ...428L..13N}. Using hydrodynamical and magnetohydrodynamical numerical simulations of accretion flows  \citet{2012ApJ...761..130Y}\citep[see also][]{Narayan2012} has shown for the first time that strong wind must exist in hot accretion flows and is responsible for the inward decrease of accretion rate obtained in \citet{1999MNRAS.310.1002S}.  This conclusion has received more and more observational evidences in recent years \citep{2013Sci...341..981W,2016Natur.533..504C,2019ApJ...871..257P,2019MNRAS.483.5614M,2021NatAs...5..928S,2022ApJ...926..209S}. It was shown that the Fermi bubbles observed by {\it Fermi}-LAT at the Galactic center may be inflated by the wind launched from the hot accretion flow in Sgr A* \citep{2014ApJ...790..109M,2015ApJ...811...37M}.

However, it is still difficult to obtained the properties of hot wind from observational data, such as mass flux and velocity, so we have to ask for the help of  theoretical studies. This is also not trivial because the accretion flow is strongly
turbulent so we need to discriminate the turbulent outflow and the real wind. If we were to calculate the mass flux of outflow at different snapshots and then perform a time average, we would overestimate the mass flux since the turbulent outflow would also be  included in this calculation. A method often adopted in the literature to filter out turbulence is to
obtain time-average physical quantities, such as mass flux and density, and then use them  to calculate the mass flux of the wind \citep[e.g.,][]{Sadowski2013}. But this will significantly underestimate the wind. This is because wind is
intrinsically instantaneous. For the same spatial region in the accretion flow, the radial velocity (or mass flux)  may be positive (i.e., it is wind) at one time but becomes negative (i.e., inflow) at another time. When we do time-average, the velocities (or mass flux) will be cancelled thus wind will be largely filtered out.

\citet{Yuan2015} solved this difficulty by using a “virtual particle
trajectory” approach (also refer to Section \ref{sec:simulation} for descriptions). Using this approach, they can trace the motion of fluid elements and distinguish turbulent outflow and ``real outflow'' (i.e., wind).  It was found that the mass flux of wind can be described by $\dot{M}_{\rm wind}=\dot{M}_{\rm BH} (r/40 r_g)$. Here $\dot{M}_{\rm BH}$ is the mass accretion rate at the black hole horizon, $r_g$ is the gravitational radius, and spherical radius $r$ measures the distance to the black hole. By taking into account both the black hole and the stellar potential of a galaxy, it was shown that the maximum $r$ is the outer boundary of the accretion flow, i.e., the Bondi radius \citep{2016ApJ...818...83B,2016ApJ...823...90B}. The large-scale dynamics of winds launched from an accretion flow is studied in \citet{2020ApJ...890...80C} and \citet{2020ApJ...890...81C}, with the inner boundary conditions mainly taken from the small-scale numerical simulation of black hole accretion. It is found that usually wind can propagate far beyond the Bondi radius. The large-scale radial profiles of velocity, density, and temperature are presented. While \citet{Yuan2015} focus only on a SANE (standard and normal evolution; \citet{Narayan2012}) around a non-spinning black hole, \citet{Yang2021} extend the \citet{Yuan2015} work to including MAD (magnetically arrested disk; \citet{Narayan2003}) and black holes with different spin values. 

When $\dot{M}\ga 2\% \dot{M}_{\rm Edd}$, the accretion is in the cold mode. This mode is further divided into the
standard thin disk \citep{Shakura1973,Pringle1981} and super-Eddington accretion \citep{Abra88,Ohsuga2005,Sadowski2014,Jiang2014}, bounded by $\dot{M}_{\rm Edd}$.  There are abundant observational evidences for the wind from a cold thin disk \citep[e.g.,][]{2003ARA&A..41..117C,2010A&A...521A..57T,2014MNRAS.443.2154T,2013MNRAS.436.2576L,2015MNRAS.451.4169G,2019NatAs...3..265H,2022SciA....8.3291H}. Many theoretical efforts have been devoted to understand the origin and acceleration of wind from a thin disk \citep[][]{2000ApJ...543..686P,2015ApJ...805...17F,2016PASJ...68...16N,2022MNRAS.513.5818W}.
For example, recently \citet{2022MNRAS.513.5818W} have performed magnetohydrodynamical numerical simulations by  considering realistic heating and cooling of the gas. They found that the model can roughly explain the column density, ionization parameter, and velocity of the observed ultra-fast outflows. Overall, the properties of wind from a cold thin disk have good observational constraint thus can be directly used in the AGN feedback study. 

In this paper, we focus on the wind launched from a super-Eddington accretion flow. Observations indicate that a significant fraction of quasars have likely gone through super-Eddington accretion during their growth \citep{2013ApJ...764...45K}, especially in their very early stage \citep{2020ARA&A..58...27I}. We are only aware of very few observations on the wind launched from super-Eddington AGNs \citep[e.g.,][]{2022ApJ...931...77D,2022A&A...668A..87V}. In the theoretical aspect, the original ``slim'' disc model, which is a one-dimensional analytical model, does not consider the wind component; but all RHD and RMHD simulations of super-Eddington accretion flow as we mention in the next paragraph show the existence of powerful wind.  

Compared to the standard thin disk, the accretion rate is much higher so radiation becomes much more important thus the physics of accretion and outflow may be significantly different from a thin disk. To study a super-Eddington accretion flow, radiation hydrodynamics simulations approach are usually adopted, although analytical work also exists \citep{cao2022arxiv}. The pioneer RHD simulation of 2D super-Eddington accretion flow was done by \citet{Eggum1988}. They found high-speed outflow near the rotation axis. Later, \citet{Ohsuga2005} developed this work to larger scale and longer simulation time. They  also found the existence of wind driven by radiation force. After that, a series of three dimensional  simulations with magnetic fields were completed, in Newtonian dynamics \citep[]{Jiang2014,Jiang2019} and  in general relativistic \citep[]{Sadowski2014,McKinney2015,  Dai2018}. Some simulation works have discussed the properties of wind using RHD \citep[e.g.,][]{2014ApJ...780...79Y,Hu2022} or RMHD simulations \citep[e.g.,][]{Sadowski2015a,Sadowski2016,Utsumi2022,Yoshioka2022arXiv}. However, the detailed properties of wind have not been systematically investigated. In addition, they usually adopt the time-averaged approach to obtain wind properties, which will underestimate the wind flux as we have explained above. At last, they usually use the value of Bernoulli parameter $Be$ to define wind and jet, which we think is not a good indicator, as we will explain in Section \ref{definition}. 

In this paper, we use trajectory approach to investigate the properties of wind and jet from a super-Eddington accretion flow. The structure of the paper is as follows. In Section \ref{sec:simulation}, we will describe our simulation setup and method. The main results of our work will be stated in Section \ref{sec:result}. We start with a definition of wind and jet in Section \ref{definition}, we then present an overview of the simulation in Section \ref{overview}. The mass flux, velocity of wind and jet, fluxes of momentum and kinetic energy and total energy are presented in Sections \ref{massflux} -- \ref{totalenergyflux}, respectively. In section \ref{accelerationmechanism}, we will conduct a force analysis to the outflow to explore whether the outflow is driven by radiation force or magnetic force. 
 We then summarize our work in Section 4. 

\section{Simulations of super-Eddington accretion flows and ``virtual particle'' trajectory approach} \label{sec:simulation}

We adopt the super-Eddington disk model ``M6a08-2'' in \citet{Thomsen2022} simulated using HARMRAD \citep{McKinney2014}. HARMRAD  is a three-dimensional general relativistic (GR) radiation magnetohydrodynamical (RMHD) code implemented using the M1 closure \citep{Levermore1984}, which includes electron scattering, thermal Comptonization, and basic free-free/bound-free processes for radiative transfer physics. Other details of the code and set up of the simulation can be found in \citet{Dai2018} and \citet{Thomsen2022}. 
In this particular model, we set the black hole mass $M_{\rm{BH}}=10^6M_{\sun}$ and  dimensionless spin $a/M=0.8$. With Thomson electron scattering opacity $\kappa_{\rm es}$, one can have the Eddington luminosity
\begin{equation}
    L_{\rm Edd}=\frac{4\pi GMc}{\kappa_{\rm es}}\approx 1.3\times10^{44}\frac{M_{\rm BH}}{10^{6}M_{\odot}}\,{\rm erg} {\rm s}^{-1}. 
	\label{eq:Ledd}
\end{equation}
The corresponding Eddington accretion rate and the nominal thin disc radiative efficiency are $\dot{M}_{\rm Edd}= L_{\rm Edd}/(\eta_{\rm NT}c^2)=1.2\times10^{24} \rm{g}\,s^{-1}$ and $\eta_{\rm{NT}} \approx 12\%$ respectively. The length unit is $\bar{L}=GM_{\rm{BH}}/c^2$, the time unit is $\bar{T}=\bar{L}/c$. The simulation sets the density unit $\bar{\rho}=1\,\rm{g}\,\rm{cm}^{-3}$, it determines the mass unit $\bar{M}=\bar{\rho}\bar{L}^{3}$. 
The simulation employs modified spherical polar coordinates with resolution of $N_r\times N_{\theta}\times N_{\varphi}=128\times 64 \times 32$. The grid in the radial direction has the inner edge located at $r_{\rm in}\approx 1.2\,r_{\rm g}$, and the outer edge located at $r_{\rm out}= 8,500\,r_{\rm g}$. The $\theta$-grid spans from $0$ to $\pi$ and has finer resolution in disc and jet region. In $\varphi$ direction, the grid is uniform from $0$ to $2\,\pi$. 
The final time of the simulation is $t_{f}$=20,000 $r_{\rm {g}}/c$. We have "inflow equilibrium radius" with "loose" criteria $r=215\,r_{\rm {g}}$ and "strict" criteria $r=111\,r_{\rm {g}}$ in time chunk t=[10,000, 20,000]$\,r_{\rm {g}}/c$ \citep{Narayan2012}. The simulation data to be used in this paper is in the time chunk t=[8,000, 12,000]$\,r_{\rm {g}}/c$.  We choose this time chunk far before the final simulation time because of the requirement of the trajectory approach. It is easy to understand that the beginning time should not be too early since otherwise the equilibrium radius would be too small. The ending time should not be too late since otherwise  the ``particles'' do not have enough time to move out of the outer boundary of simulation domain, which is required in our trajectory approach to judge whether the outflow is turbulence or real wind. The corresponding ``inflow equilibrium radii'' with "loose" and ``strict'' criteria are $r=134\,r_{\rm {g}}$ and $r=80\,r_{\rm {g}}$, respectively.

To ensure the simulation has well resolved the magneto-rotational instability (MRI), we check the MRI fastest growing wavelength,
\begin{equation} \label{MRI}
    \lambda_{x, \rm MRI} \approx 2\pi \frac{|v_{x,\rm A}|}{|\Omega|}
\end{equation}
for $x=\theta$ or $\phi$ \citep{Hawley.1995}. Here, $|v_{x, \rm A}| = \sqrt{b_x b^x/(b^2+\rho+u_g+p_g)}$ is the Alfv\'en speed in the x direction. 
This allows us to check the number of grid cells per fastest-growing MRI mode in $\theta$ and $\phi$ directions, namely the MRI quality factor, 
\begin{equation} \label{Q_MRI}
    Q_{x, \rm MRI} = \frac{\lambda_{x, \rm MRI}}{\Delta_{x}}
\end{equation}
for $x=\theta$ or $\phi$, where $\Delta_{x}$ is the $x$-grid cell length. If $Q_{\rm x, MRI}>6$, the MRI is considered as sufficiently resolved. In the time window of $t=[8000, 12,000] r_g/c$, we find the values of the $t$-$\theta$-$\phi$-averaged $Q_{\rm \theta, MRI}$ and $Q_{\rm \phi, MRI}$ have passed the criteria within the loose inflow equilibrium radii ($r=134 r_g$), meaning that the MRI is well resolved in $\theta$ and $\phi$ directions in the simulated accretion inflow.


After we obtain the simulation results, we then use the ``virtual particle trajectory'' approach to analyze the data. Readers are referred to \citet{Yuan2015} for the details of this approach or Section 2.3 in \citet{Yang2021} for a brief description. Very briefly speaking, we uniformly put $64\times 32$ virtual test particles at a certain radius with different $\theta$ and $\phi$ and obtained their trajectories using our simulation data.  From the trajectories, we can judge whether they belong to real outflow (i.e., wind) or turbulence. The criteria is whether the particle can keep moving outward and never change its direction to cross that radius. We then be able to calculate the properties of wind. The mass flux of wind is calculated by summing up the corresponding mass flux carried test particles whose trajectories belong to the real outflow (refer to eq. \ref{eq:mdots}). In our previous works \citep{Yuan2015,Yang2021}, we have compared the mass flux of wind obtained by this method and that obtained by the method of, e.g., \citet{Sadowski2013}. We find that the former is larger than the latter by a factor of 2-6, depending on radius.

It is not clear in nature whether the super-Eddington accretion flows in AGNs belongs to an MAD or SANE. Current studies to blazars indicate that the accretion flows in them  are likely MAD  \citep{2014Natur.510..126Z,2014Natur.515..376G,2015MNRAS.451..927Z}. For low-luminosity AGN M~87 and Sgr A*, the EHT has provided evidences that they are likely also MAD \citep{2021ApJ...910L..13E,2022ApJ...930L..16E}.  The MAD nature of M~87 is confirmed later by comparing the  observed values of rotation measure along its jet with theoretical predictions \citep{2022ApJ...924..124Y}.  Because whether the accretion flow is MAD or SANE is mainly determined by the coherence of the magnetic field configuration in the accreted  matter \citep{Narayan2003}, which is not expected to correlate with the accretion rate, we assume that super-Eddington accretion flows in many AGNs are in this mode and focus on MAD in the present work.  
We focus on a relatively high black hole spin of $a=0.8$. This is because, on one hand, we are interested in the comparison between jet and wind; on the other hand, the measured spin values of most black holes seem to be large (see the combination of the measured supermassive black hole spin values so far in  \citet{2019ApJ...873..101Z} and the review by \cite{2021ARA&A..59..117R}).

\section{Results} \label{sec:result}

\subsection{Definitions of jet and wind in simulations}
\label{definition}

Before we present our results, we first introduce our definitions of jet and wind. First, both of them  must be ``real outflow'', i.e, they must be outflows that keep moving outward to large enough distance. ``Real outflow'' is discriminated from ``turbulent outflows''; the latter changes their direction of motion and returns to the black hole, as indicated by their trajectories (refer to Figure 2 in \citet{Yuan2015} and relevant descriptions). The next question is how to discriminate jet and wind. We follow the approach presented in \citet{Yang2021}. That is, the boundary between jet and wind is represented by the magnetic field line whose foot point is rooted at the black hole ergosphere  with $\theta=90^{\circ}$, i.e., the boundary between the black hole ergosphere and  the accretion flow. We note that this definition coincides with other notions of the jet based on velocity and magnetization, as seen even in early two-dimensional simulations \citep[e.g.,][]{2004ApJ...611..977M}. Under this definition, all magnetic field lines in the jet region are anchored to the black hole ergosphere and thus can extract the spin energy of the black hole via the Blandford \& Znajek mechanism \citep{Blandford1977Znajek,Tchekhovskoy2011} to power the jet. Real outflows outside of this boundary are powered by the accretion flow and we call them wind. 

We further explain the physical basis of the above definition as follows. In the observational side, jet is well defined from their prominent morphology. But theoretically it is always not clear how the simulated jet is mapped with the observed one. The problem becomes more difficult given that it has been proposed both the rotating accretion flow and the spinning black hole are able to power jets via large-scale magnetic field, called ``disk-jet'' and ``BZ-jet'' respectively \citep{Blandford1977Znajek,1982MNRAS.199..883B}.  

Most recently, by employing general relativity MHD simulation of black hole accretion and jet formation and considering electron acceleration by magnetic reconnection, \citet{Yang2022} have calculated the predicted image of the BZ-jet and compared the results to the millimeter observation of the jet in M~87. They find that the BZ-jet can well reproduce the observed jet morphology, including the jet width and limb-brightening feature, while the disk-jet is ruled out. This result convincingly shows  that the observed jet must correspond to the BZ-jet. Although this study is for the jet launched from a hot accretion flow, in this paper, we assume that the jet launched from a super-Eddington accretion flow is the same and we adopt the same definition of jet for super-Eddington accretion flow. 

Such definitions of jet and wind are different from those adopted in some papers. In those works, outflow (wind and jet) are often defined as those outflows satisfying $u^{r}>0$ and  Bernoulli parameter $Be>0$. Here $Be$ is defined as \citep{Sadowski2013,Sadowski2015b},
\begin{equation}
    Be=-\frac{T^{r}_{t}+R^{r}_{t}+\rho u^{r}}{\rho u^{r}}.
	\label{eq:Be}
\end{equation}
The first term denotes the energy-momentum tensor (including both the fluid and magnetic field), the second term the energy-momentum tensor of the radiation field. When $Be>0.05$, they call the outflow jet; when $0<Be<0.05$, it is called wind. The problem with this definition is that the Bernoulli parameter $Be$ is usually not a constant due to the turbulent nature of the accretion flow, as shown by tracing the trajectories of virtual Lagrangian test particles and calculating their values of $Be$ \citep{Yuan2015}. It is found that some outflows  with initial $Be<0$ gain energy and finally escape from the black hole. In agreement with \citet{Yuan2015}, recently \citet{Yoshioka2022arXiv} have also found that the value of $Be$ along the streamline of wind could change and concluded that the sign of $Be$ is not a correct indicator of wind or jet. 

\subsection{Overview of simulation results}
\label{overview}

\begin{figure*}
   \includegraphics[width=1.0\linewidth]{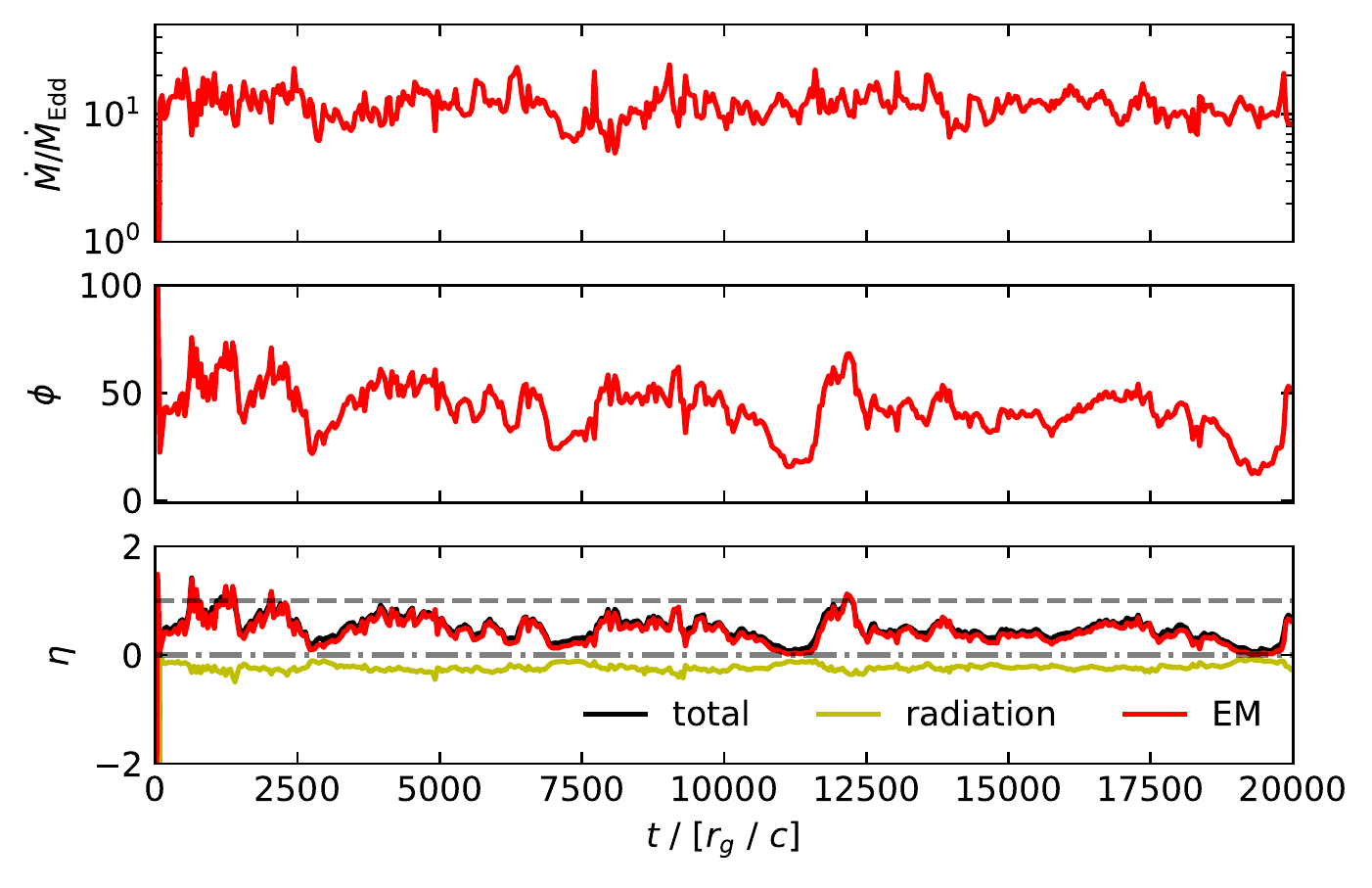}
   \caption{The time evolution of different quantities measured at the black hole horizon. From top to bottom, they are the mass flux of accretion flow in unit of $\dot{M}_{\rm Edd}$, the magnetic flux that threads the hemisphere of the black hole horizon, and the efficiencies of the total energy flux (black), radiation energy flux (yellow) and Poynting flux (EM; red) respectively.}
   \label{fig:t-mdot}
\end{figure*}

\begin{figure*}
   \includegraphics[width=0.31\linewidth]{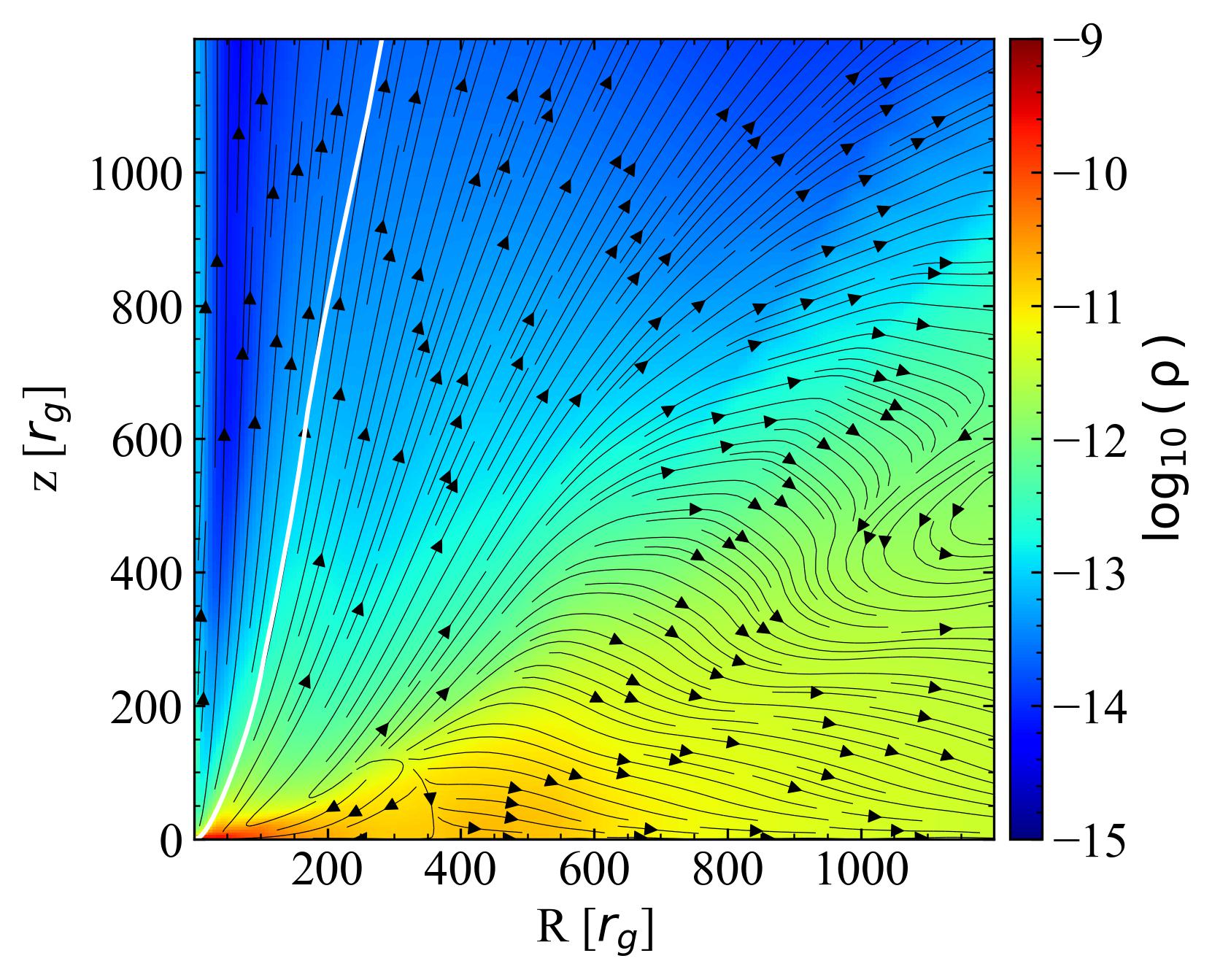}
   \includegraphics[width=0.31\linewidth]{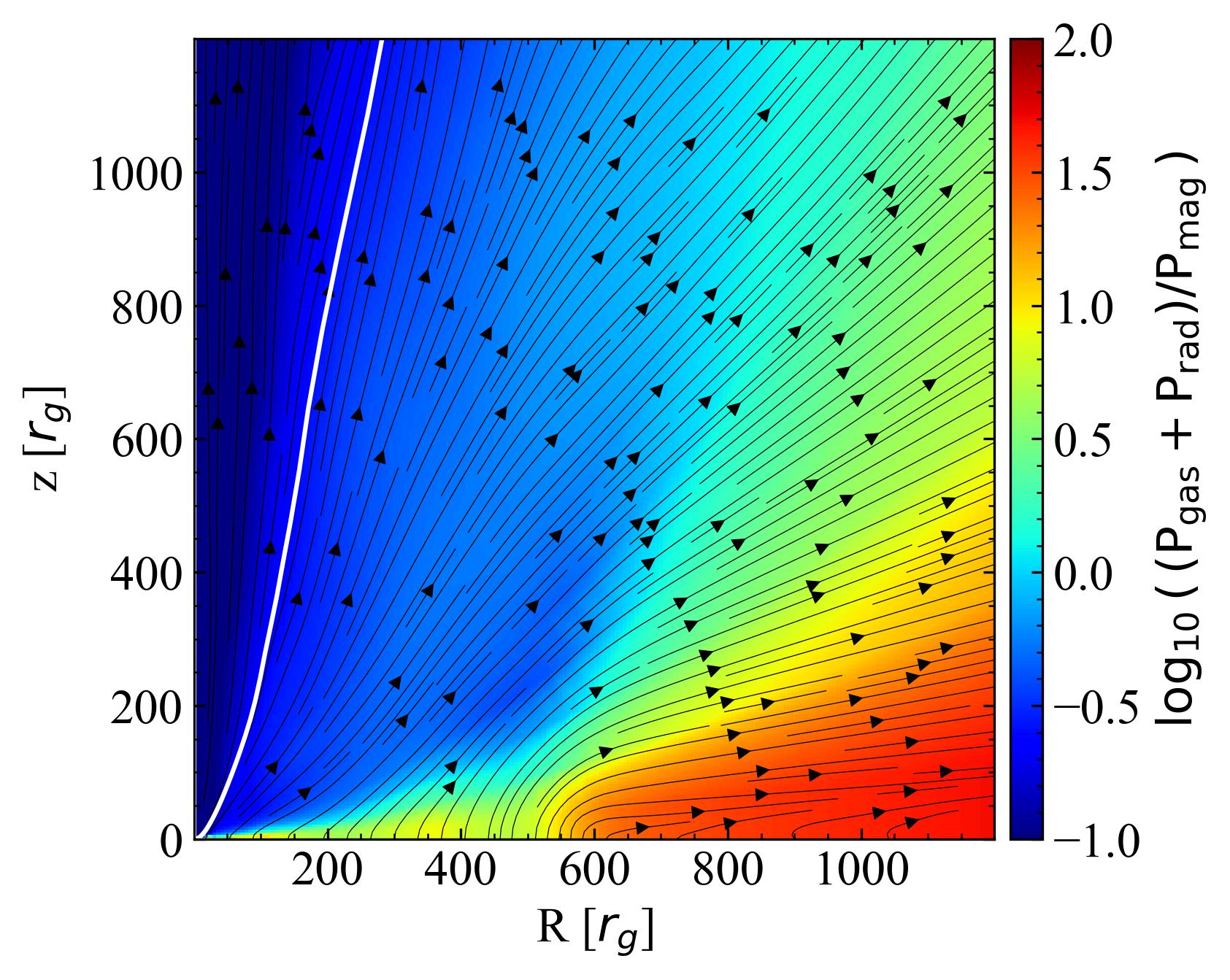}
   \quad\includegraphics[width=0.31\linewidth]{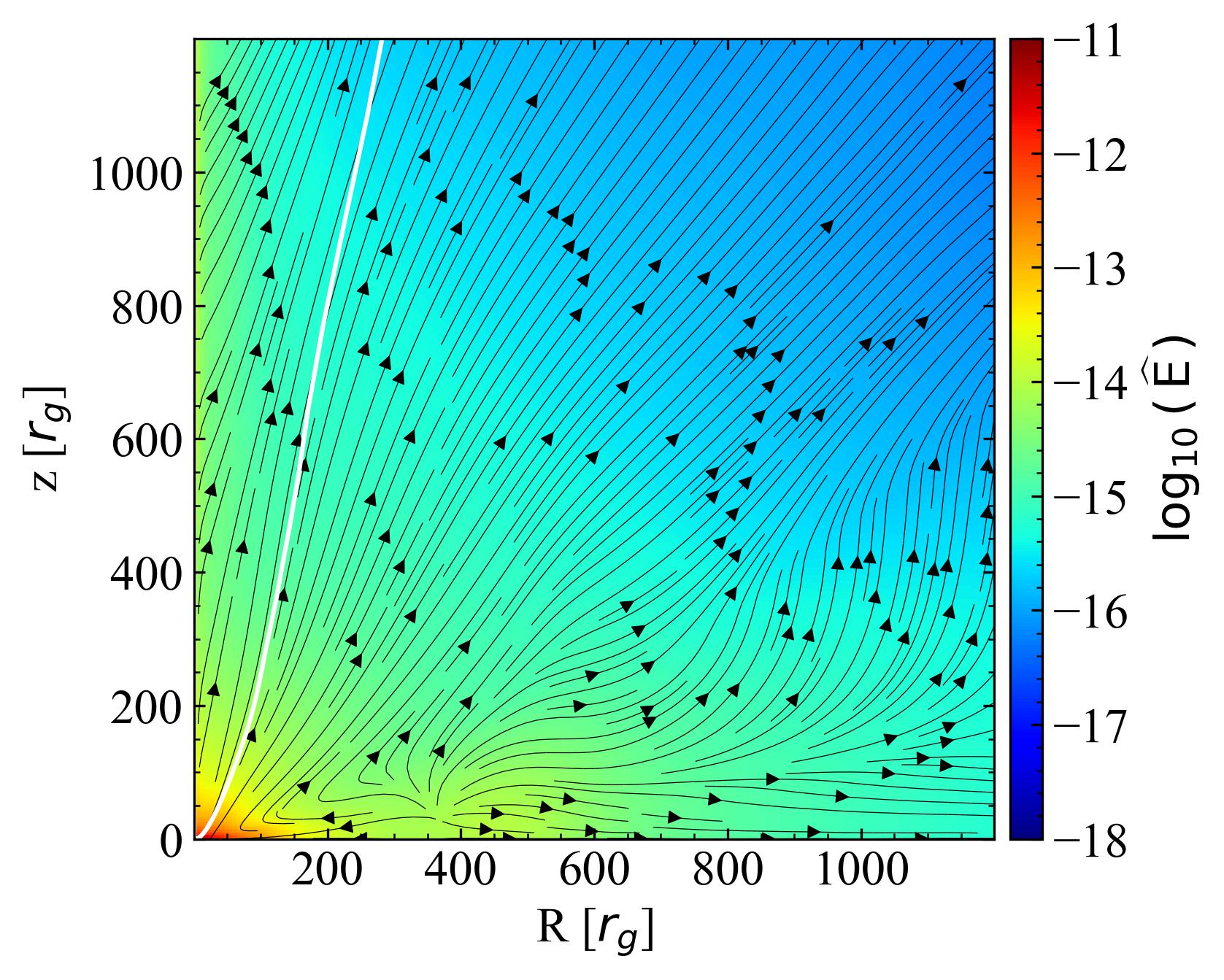}
   \caption{The time-averaged two-dimensional distributions of density (left), plasma $\beta=(P_{\rm{gas}}+P_{\rm{rad}})/P_{\rm {mag}}$ (middle), and the radiation energy density $\widehat{E}$ (right). From left to right, the the black solid lines denote the streamline of velocity field,  magnetic field lines, and the radiation field stream lines, respectively. The white solid line in all panels denote the boundary between the BZ-jet and wind.}
   \label{fig:fieldline}
\end{figure*}

 \begin{figure*}
     \centering
     \includegraphics[width=0.48\linewidth]{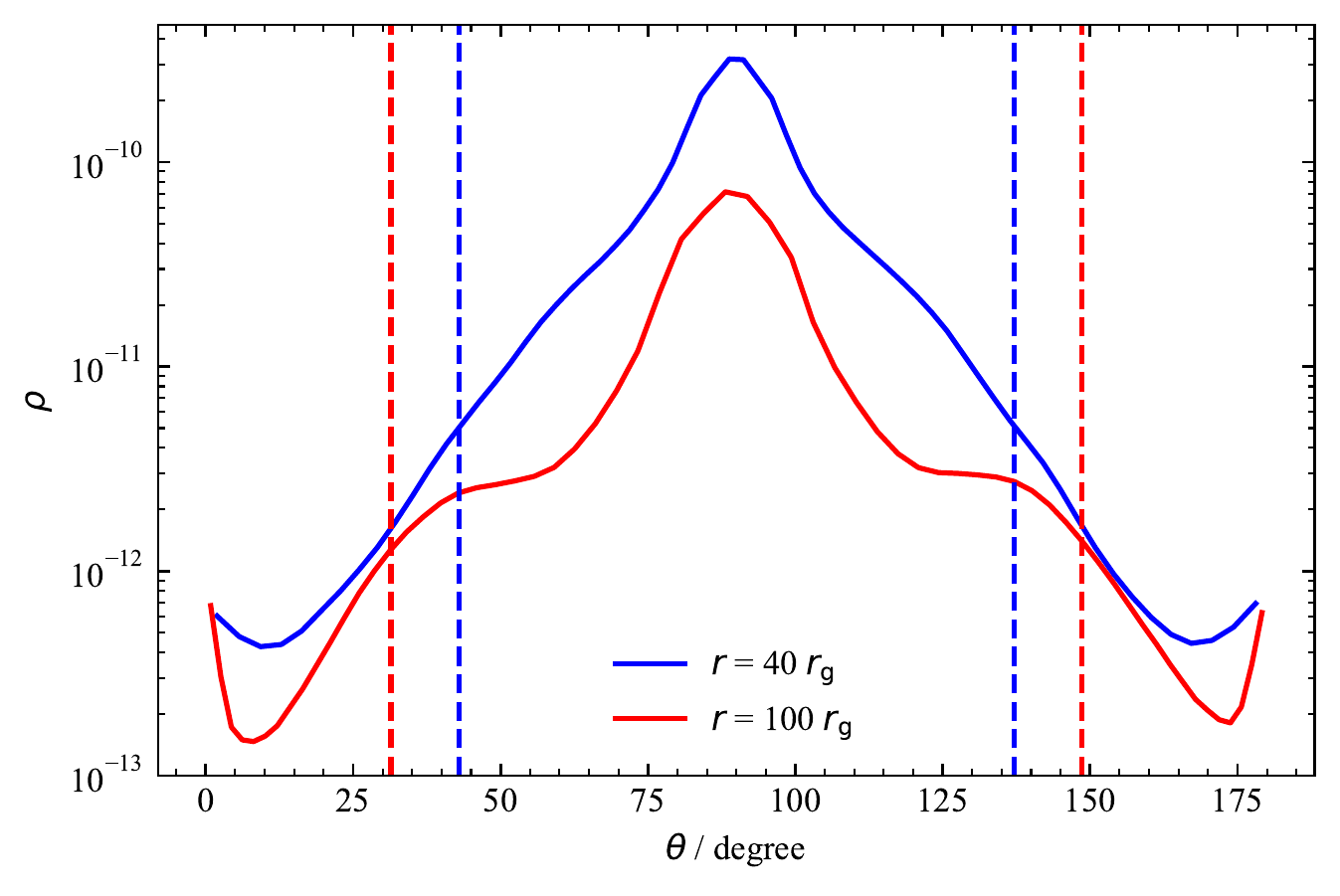}\quad\includegraphics[width=0.48\linewidth]{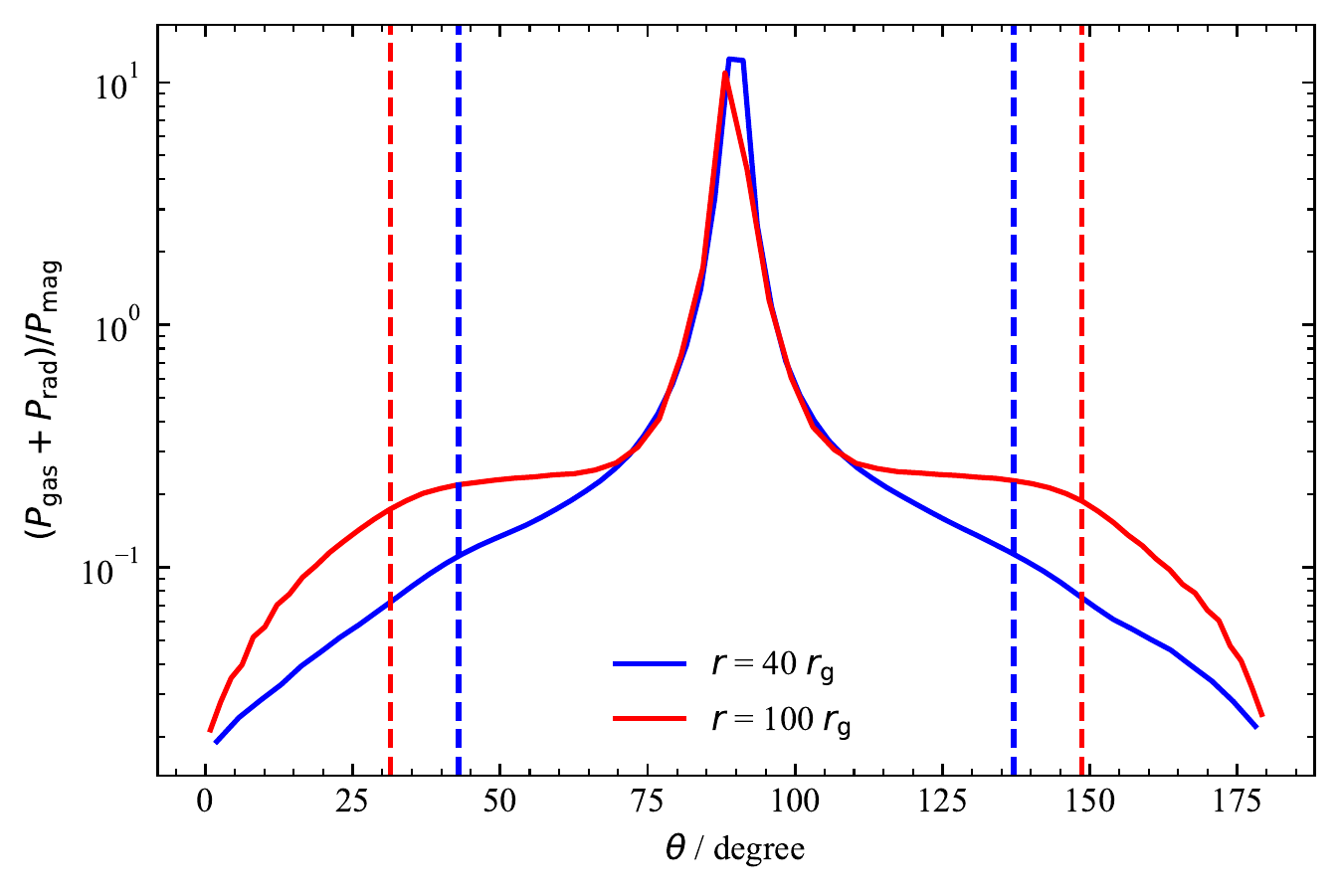}
     \includegraphics[width=0.48\linewidth]{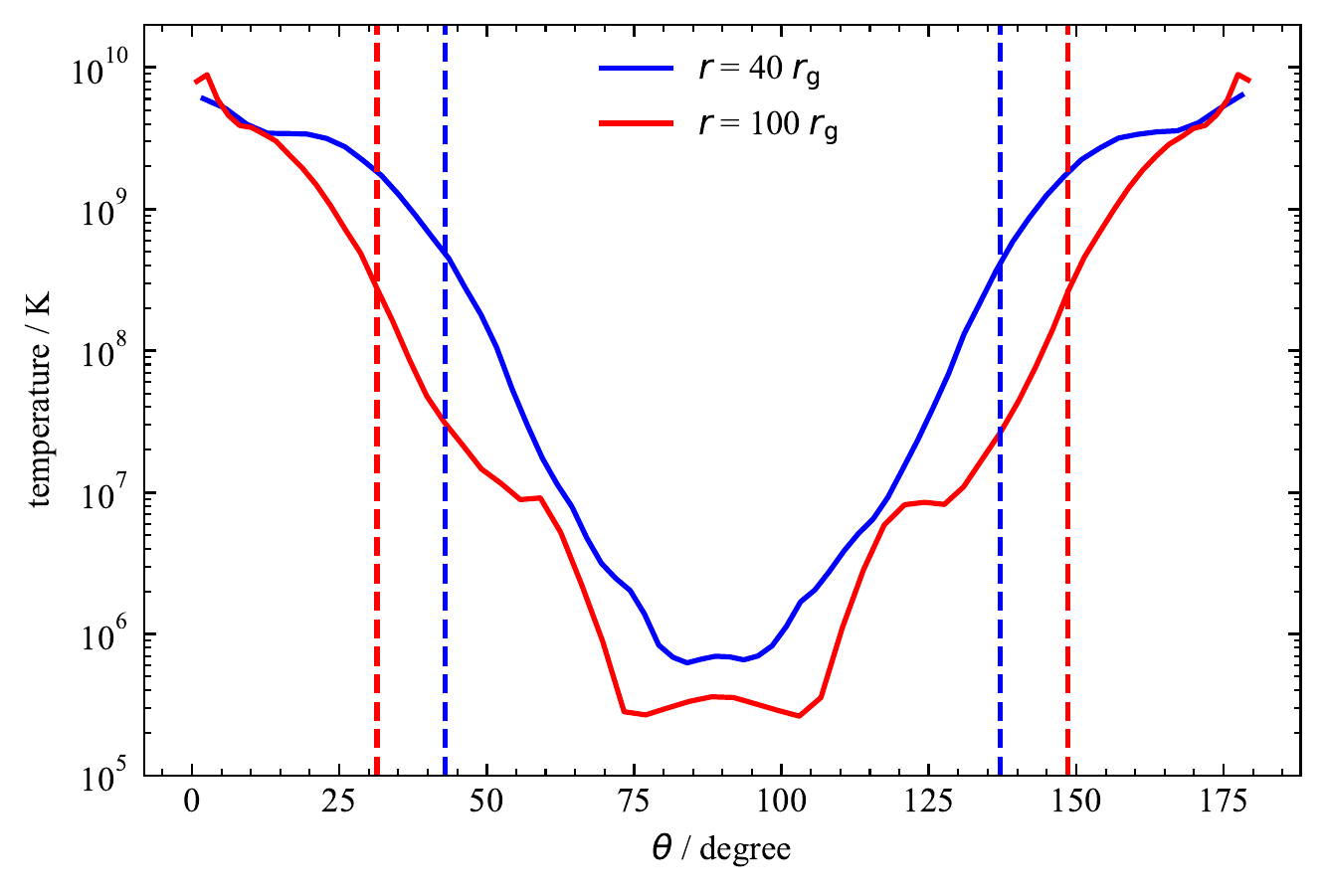}\quad\includegraphics[width=0.48\linewidth]{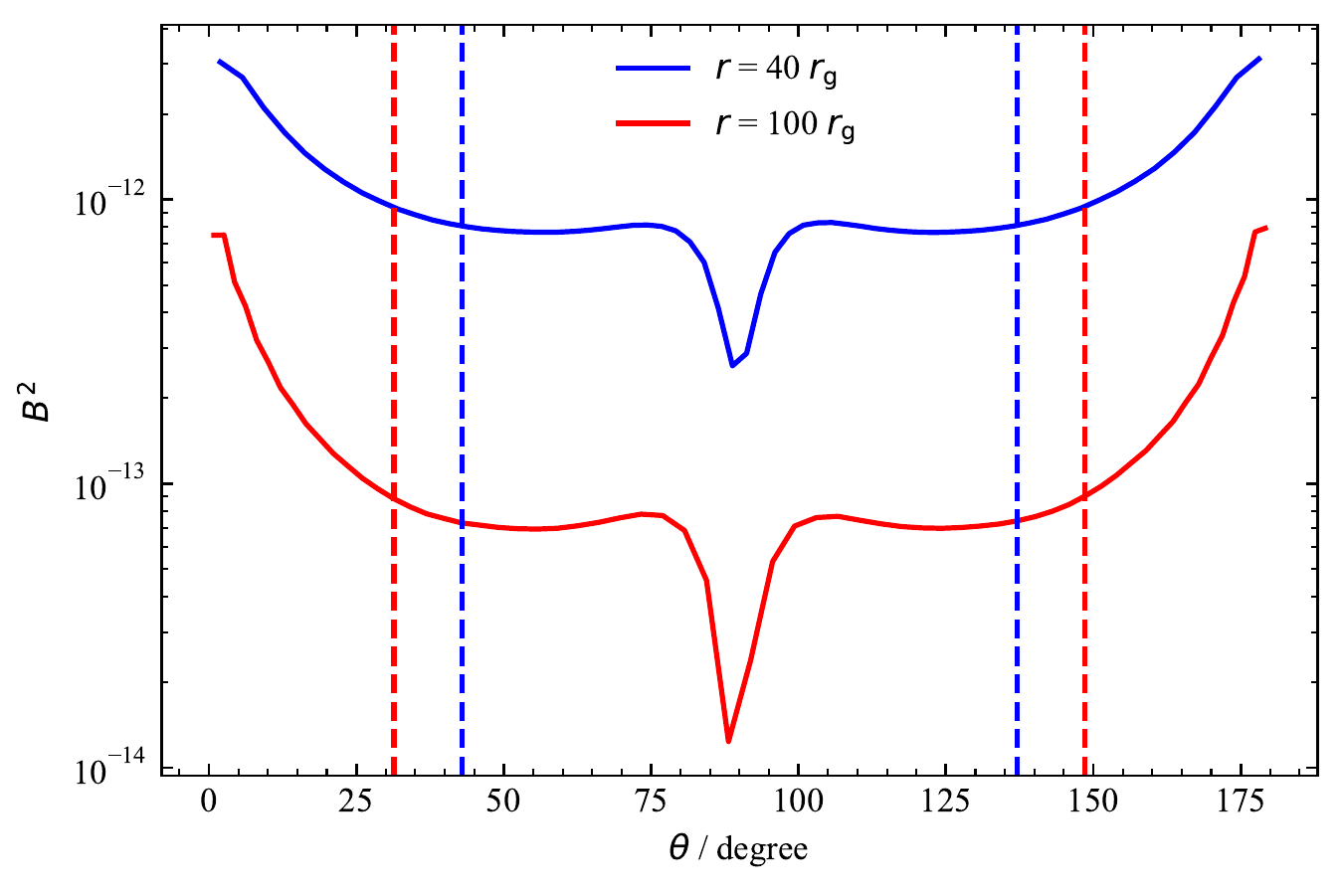}
     \caption{The values of some quantities averaged over $\varphi$ and $t$ as a function of $\theta$ at $r=40\,r_{\rm g}$ (blue) and $r=100\,r_{\rm g}$ (red). The vertical dashed line denotes the boundary between the BZ-jet and wind.}
     \label{fig:fdensity}
 \end{figure*}

\begin{figure*}
     \centering
   \includegraphics[width=0.8\linewidth]{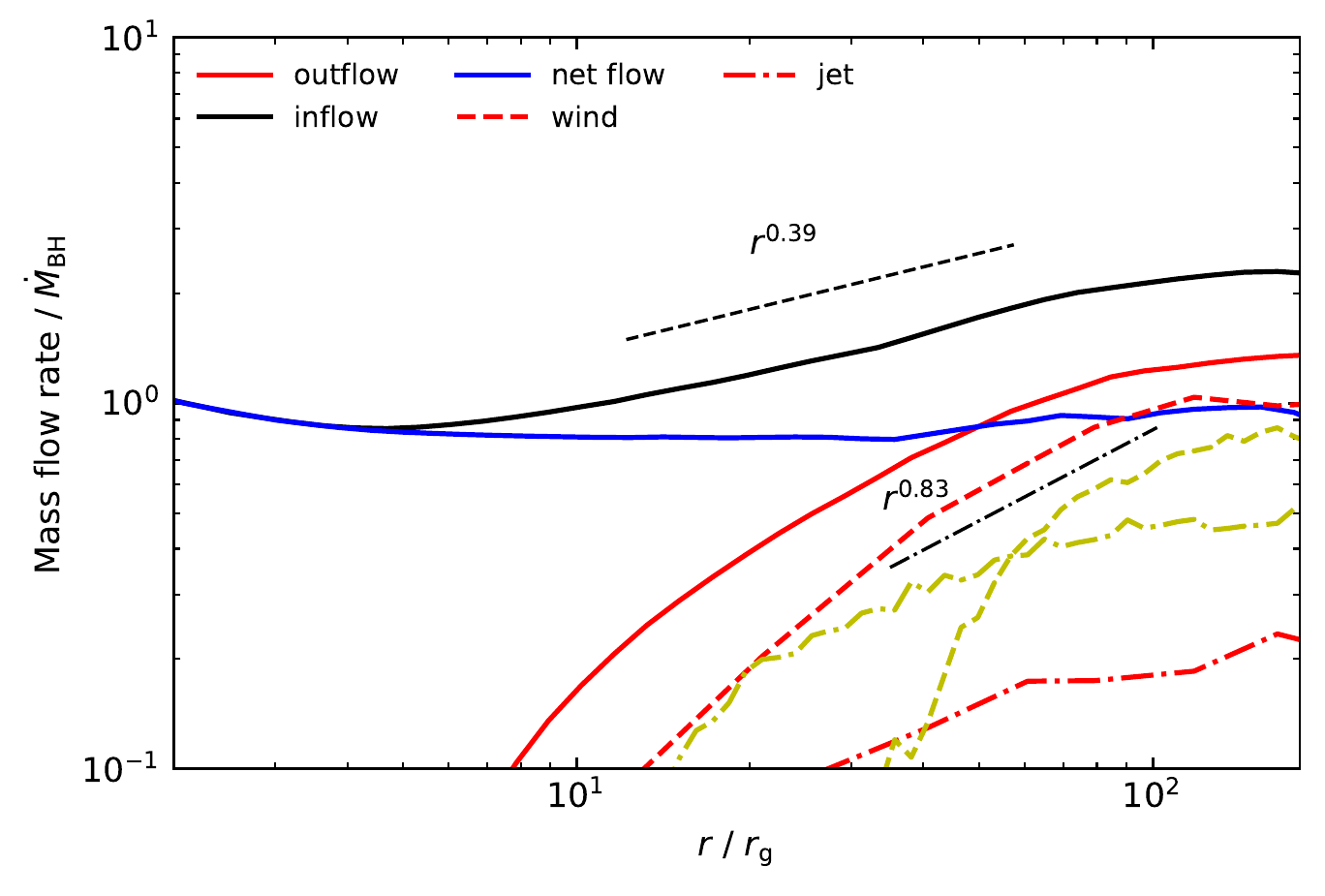}
   \caption{The radial profiles of inflow rate (red solid line), outflow rate (black solid line), net rate (blue solid line), mass flux of wind (red dashed line), and mass flux of jet (red dot-dashed line). The yellow dashed line and yellow dot-dashed line are the mass fluxes of wind and jet  obtained by the time-averaged streamline method. All values have been normalized by the mass accretion rate at the black hole horizon $\dot{M}_{\rm BH}$.  }
   \label{fig:wind}
\end{figure*}

The mass accretion rate at radius $r$ is calculated by
\begin{equation}
    \dot{M}(r)=-\int_{0}^{2\pi}\int_{0}^{\pi}{\rho}u^{r}\sqrt{-g}d\theta d\varphi.
	\label{eq:Mdot}
\end{equation}
 The energy extraction efficiency, or the energy outflow efficiency, $\eta$, is defined as the energy return rate to infinity divided by the rest-mass accretion power \citep{Yang2021}:
\begin{equation}
 \centering
 \eta= \frac{|\dot{M}_{BH}|c^2-\int_{}^{} (T^r_t+R^r_t )d A_{\theta\varphi}}{|\dot{M}_{\rm{BH}}|c^2}.
 \label{eq:eta}
\end{equation}
Here $T^{\mu}_{\nu}$ represents the total energy flux (as measured at infinity) transported by the fluid and the magnetic field, including the rest-mass energy of the gas, $R^{\mu}_{\nu}$ is the radiation energy-stress tensor, $\sqrt{-g}$ is the metric determinant, $\dot{M}_{\rm BH}$ is the mass accretion rate at black hole horizon. The magnetic field flux that threads the hemisphere of the black hole event horizon is calculated as follows,
\begin{equation}
    \Phi(r)=\frac{1}{2}\int_{0}^{2\pi}\int_{0}^{\pi}\sqrt{4\pi}|B^{r}| \sqrt{-g} d\theta d\varphi.
	\label{eq:bflux}
\end{equation}
Following \citet{Tchekhovskoy2011}, a dimensionless magnetic flux normalized with the mass accretion rate is then defined as
\begin{equation}
    \phi(r)=\frac{\Phi}{\sqrt{\dot{M}_{\rm BH}c}r_{\rm g}}.
	\label{eq:phi}
\end{equation}

Figure~\ref{fig:t-mdot} shows the time evolution of the mass accretion rate, magnetic flux and total energy efficiency measured at the black hole event horizon. We can see that the mass accretion rate is roughly 10\,$\dot{M}_{\rm{Edd}}$, the average mass accretion rate within $t=[8{,}000, 12{,}000]\,r_{\rm g}/c$ is 11.5\,$\dot{M}_{\rm{Edd}}$. The dimensionless magnetic flux fluctuates around 50, so the accretion flow has reached an MAD state. The total energy efficiency is larger than 100\% at some time. 
The power of the BZ-jet and the corresponding efficiency are calculated by,
\begin{equation}
    P^{\rm BZ}=\frac{\kappa}{4\pi}{\Omega}^2_{\rm H}{\Phi}^2_{\rm BH}f(\Omega_{\rm H}),
	\label{eq:Pbzn}
\end{equation}
\begin{equation}
    \eta^{\rm BZ}=\frac{P^{\rm BZ}}{\dot{M}_{\rm BH}c^2},
	\label{eq:etaPbz}
\end{equation}
respectively. Here $\kappa$ is a numerical constant that depends on the geometry of the magnetic field and we adopt $\kappa=0.044$ in this paper, $\Omega_{\rm H}=ac/2r_{\rm H}$ is the angular velocity of the black hole horizon, and $f(\Omega_{\rm H})$ is a modifying factor for high spin $a$, which is $f(\Omega_{\rm H})\approx1+1.38({\Omega_{\rm H}r_{\rm g}}/{c})^2-9.2({\Omega_{\rm H}r_{\rm g}}/{c})^4$ \citep{Tchekhovskoy2010}.
The calculated BZ-jet efficiency $\eta_{\rm{BZ}}=38.8\%$, while the total efficiency calculated from Equation (\ref{eq:eta}) is $\eta=47.6\%$. 

We have also calculated the efficiencies of the Poynting flux and radiation flux at the black hole horizon defined as 
\begin{equation}
 \centering
 \eta_{\rm{P}}= \frac{-\int_{}^{}(b^2 u^r u_t-b^r b_t d) A_{\theta\varphi}}{|\dot{M}_{\rm{BH}}|c^2},
 \label{eq:p}
\end{equation}

\begin{equation}
 \centering
 \eta_{\rm{rad}}= \frac{-\int_{}^{} R^r_t d A_{\theta\varphi}}{|\dot{M}_{\rm{BH}}|c^2}.
 \label{eq:rad}
\end{equation}
 We found $\eta_{\rm{P}}=40.6\%$. Combining with $\eta=47.6\%$, this means that the Poynting flux contributes a fraction of $40.6/47.6=85\%$ of the total energy flux, i.e., it is the dominant energy flux at the horizon.  The radiation flux is $\eta_{\rm{rad}}=-23.6\%<0$, it means that the net radiation flux flows into the black hole horizon (refer to Figure 20 in \citet{McKinney2014} and Figure 6 in  \citet{Sadowski2016}),  which is because the photons are trapped and dragged (advected) with the inflow into the horizon. The radiation luminosity is almost a constant beyond $\sim 200 r_g$, which is about $4.8\% \, \dot{M}_{\rm BH}c^2$ \citep{Thomsen2022}.

 Figure~\ref{fig:fieldline} shows the spatial distribution of density $\rho$, plasma $\beta$, and radiation energy density $\widehat{E}$  averaged in $\phi$ direction and within the time chunk $t=[8{,}000\,, 12{,}000]\,r_{\rm{g}}/c$. The radiation energy density reaches its largest value near the spin axis of black hole and on the equatorial plane. The radiation flux is inward at small radii of the equatorial plane, while in the wind and jet regions it is almost always outward. The white solid line in the figure denotes the boundary between the BZ-jet and wind. 
 
Figure~\ref{fig:fdensity} shows in a more quantitative way various quantities of the flow averaged over $\varphi$ and time as a function of $\theta$ at $40\,r_{\rm g}$ and $100\,r_{\rm g}$, including density, the ratio of the gas pressure plus radiation pressure to magnetic pressure $\beta$, temperature, and $B^2$. Compared with the results of hot accretion flows   \citep{Yang2021}, it is interesting to note that, in both cases, both the density and $B^2$ cover roughly three orders of magnitude from the equatorial plane to the rotation axis, while the gas temperature covers nearly five orders of magnitude in the present super-Eddington case, much larger than the value of three orders of magnitude in the case of hot accretion flows. 

\subsection{Mass fluxes of wind and jet}
\label{massflux}

To calculate the mass flux of the wind at a given time $t_0$ and radius $r$, we first put test particles at that radius $r$ with different $\theta$ and $\varphi$ and obtain their trajectories starting at $t_0$. The mass flux of wind is calculated by summing up the corresponding mass flux carried by test particles whose trajectories belong to the ``real outflow'' \citep{Yuan2015} using the following formula:
\begin{equation}
    \dot{M}_{\rm wind (jet)}( r)=\sum_{i}\rho_i(r )u^r_i( r)\sqrt{-g}\delta\theta_i\delta\varphi_i.
	\label{eq:mdots}
\end{equation}
Here the subscript represents different test particles, $\rho_i$ and $u^r_i$ are the mass density and the radial contravariant component of the 4-velocity at the location where the test particle ``$i$'' originates, and $\delta\theta_i$ and $\delta\varphi_i$ are the ranges of $\theta$ and $\varphi$ the particles occupy. 

 In our calculations, we usually choose 10 different initial times $t_0$ to obtain the trajectories of these test particles and the mass flux corresponding to each choice of $t_0$. We then do time-average of these 10 groups to obtained the averaged mass flux. These 10 initial times are uniformly selected after the simulation has reached steady state. Because the calculation of trajectory requires simulation data after time $t_0$, and it needs ``test particles'' to move away from their initial location far enough to distinguish the real outflow from turbulence, the initial time $t_0$ should not be too late. The time intervals we choose for the time-average is $t_0=8{,}000$ to $12{,}000$. To calculate the wind properties (such as mass flux, velocity, and energy and momentum fluxes) as a function of radius, we have chosen a series of initial radii, namely 
 {\bf $r=10\,r_{\rm g}$, $20\,r_{\rm g}$, $40\,r_{\rm g}$, $60\,r_{\rm g}$, $80\,r_{\rm g}$, $120\,r_{\rm g}$, $160\,r_{\rm g}$, $200\,r_{\rm g}$}. The initial position of the ``test particle'' in $\theta$ and $\varphi$ directions are consistent with our simulation grid, i.e., 64 grid points in the $\theta$ direction $[0,\pi]$ and 32 grid points in the $\varphi$ direction $[0, 2\pi]$. 
 
The radial profiles of the calculated mass fluxes of  wind and jet are shown in Figure~\ref{fig:wind}.  The black, red, and blue solid lines show the inflow,  outflow, and net rates, respectively. The inflow and outflow rates are calculated using Equation~(\ref{eq:Mdot}) and the net rate is the difference between them. Note that here the outflow rate $\dot{M}_{\rm out}$($=\int{\rho \, {\rm max}\{u^r,0\}  d A_{\theta\varphi} }$) includes both real outflow (jet and wind) and the turbulent outflow. The red dashed and dot-dashed lines denote the mass fluxes of  wind and jet respectively, obtained using our trajectory approach. 
Beyond several $r_g$, the radial profiles of the inflow rate can be fitted by 
\begin{equation}
    \dot{M}_{\rm in}(r )=\left(\frac{r}{3\,r_{\rm s}}\right)^{0.39}\dot{M}_{\rm BH},
	\label{eq:inflow}
\end{equation}
The radial profiles of the mass flux of wind can be fitted by
\begin{equation}
    \dot{M}_{\rm w}(r )=\left(\frac{r}{45\,r_{\rm s}}\right)^{0.83}\dot{M}_{\rm BH},
	\label{eq:wind}
\end{equation}
Here $\dot{M}_{\rm BH}$ is the mass accretion rate at the black hole horizon. When applying the above formulae to calculate the mass flux of wind, one question one may ask is what is the largest value of $r$  we should adopt? As we have stated in the introduction, this question has been investigated in the case of hot accretion flow wind, and it was found that the largest $r$  is the outer boundary of the accretion flow, which is ``truncation radius'' if the accretion flow is a hybrid thin disk plus hot accretion flwo structure \citep{Yuan2014Narayan} or the Bondi radius if the hot accretion flow is not truncated. We speculate that it is similar to the present super-Eddington case, i.e., wind cannot be produce beyond the Bondi radius, although of course the wind produced within the Bondi radius can propagate far beyond the Bondi radius.

The super-Eddington accretion flow has some similarities with the hot accretion flow. Both are advection-dominated and geometrically thick, and in both cases winds are strong in unit of their respective mass accretion rates. It is thus useful to compare their wind properties. Compare Equation (\ref{eq:wind}) with the corresponding wind flux formula in the case of a hot accretion flow \citep{Yang2021}, we find that the current ``denominator'' (i.e., 45$r_g$) is somewhat larger than the corresponding value for an MAD hot accretion flow around a rapidly spinning black hole, where it is $30\,r_{\rm g}$; while the index (i.e., $0.83$)  is a bit smaller than that of hot accretion flow, which is  $1.26$. These two facts imply that the mass flux of wind in the case of super-Eddington accretion is somewhat weaker than that in the hot accretion flow in unit of their respective accretion rates at the black hole horizon, although the absolute mass flux of wind in the case of super-Eddington accretion is much stronger than the case of hot accretion flows. This may be  because the gas temperature of a super-Eddington flow is typically two orders of magnitude lower than that of a hot accretion flow, as we can see by comparing Figure \ref{fig:fdensity} in the present paper and Figure 4 in \citet{Yang2021}. In fact, the thermal energy of the gas in a hot accretion flow is comparable to its gravitational energy; consequently, its Bernoulli parameter is close to zero thus the gas can easily escape from the black hole and forms wind \citep{1994ApJ...428L..13N,Yuan2014Narayan}. In the case of super-Eddington accretion flow, the thermal energy of the gas is much smaller than its gravitational energy. The radiation pressure is high and potentially  can help the acceleration of wind, but we find that in most of the region the radiation energy density is still over one order of magnitude smaller than the gravitational energy. This may be the reason why the wind in the current super-Eddington case is somewhat weaker than that in hot accretion flows.

The mass flux of jet is very small at small radii. It rapidly increases and roughly saturates at $\sim 60 r_g$. The saturated value is 
\begin{equation}
\dot{M}_{\rm jet} \approx 0.23 \dot{M}_{\rm BH}.
\label{jetflux}
    \end{equation} 
This result is again similar to the case of  hot accretion flows \citep{Yang2021}. In that case, the mass flux of jet saturates at $20-100r_g$, depending on the magnetization of the accretion flow (``MAD'' or ``SANE''). But the saturated value in the case of hot accretion flow in unit of $\dot{M}_{\rm BH}$ is higher, which is $0.5-0.8 \dot{M}_{\rm BH}$. Again, the main reason for such a discrepancy is that, the temperature of a super-Eddington accretion flow is much lower than that of a hot accretion flow. 

\begin{figure}
 
   \includegraphics[width=1.0\linewidth]{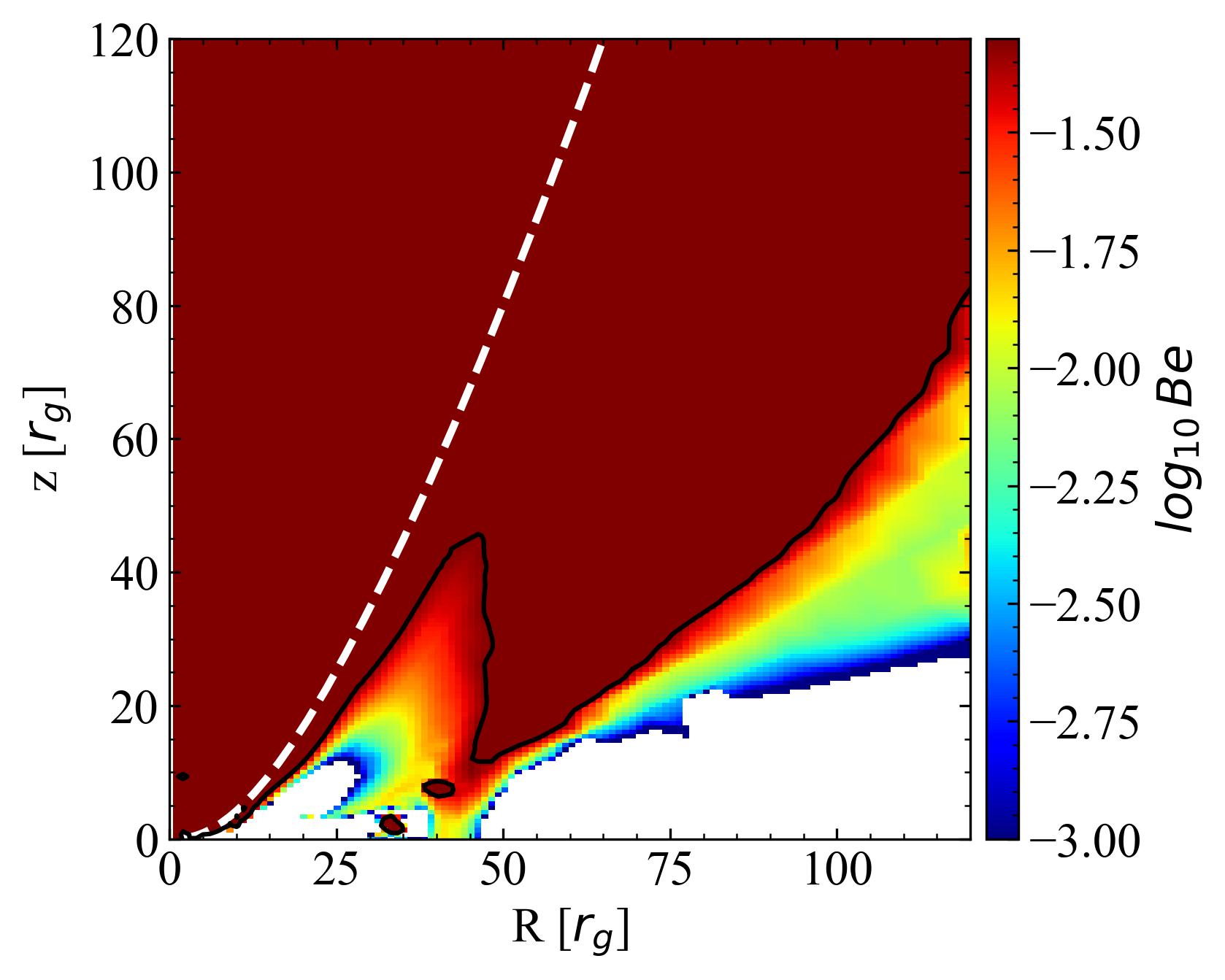}
   \caption{ The distribution of the value of $Be$ at t=8,000 $\,r_{\rm {g}}/c$.  The black solid line is the contour of $Be=0.05$. The white area denotes $Be<0$. The white dashed line denotes the boundary between BZ-jet and wind. }
   \label{fig:Be}
\end{figure}

 For comparison purpose, we have also used the ``time average'' method widely adopted in literature \citep[e.g.,][]{Sadowski2013} to calculate the mass flux of the wind and jet. To do this, we first time average the physical quantities such as density, velocity and mass flux ($\rho u^{r}$) in the time from 8,000 $r_{\rm g}/c$ to 12,000 $r_{\rm g}/c$ and in the $\varphi$ direction.  We  use the definitions of wind and jet as described in Section \ref{definition}, with the boundary between wind and jet being $Be=0.05$. The results for the mass flux of the wind and jet using this method are shown by the yellow dot-dashed  and dotted lines in Figure~\ref{fig:wind}. We find that the mass flux of the wind obtained by the streamline method is smaller than that obtained by the trajectories method by a factor of $2-6$, depending on radius. Such a difference is similar to the case of wind from a hot accretion flow \citep{Yuan2015,Yang2021}. We speculate that the reason for the difference are two-fold. Wind is intrinsically instantaneous. In the streamline method, by doing time-average, some wind will be ``filtered out''. This is possibly the dominant reason. In addition, the required range of $0\la Be \la 0.05$ for wind is a bit too narrow, since we find outflows beyond this range can also escape as wind.  For the mass flux of jet, from the figure we can see that the value obtained by the streamline method is larger than that obtained by our trajectory method by a factor of $\ga 2$. The main reason for this may be the difference of the definition of jet. As can be seen in Figure~\ref{fig:Be}, which shows the distribution of time-snapshot $Be$, the opening angle of the jet defined by the streamline method is nearly two times larger than that of BZ-jet. 

\subsection{Poloidal and toroidal velocities of wind and jet}
\label{velocity}

\begin{figure}
\includegraphics[width=0.9\linewidth]{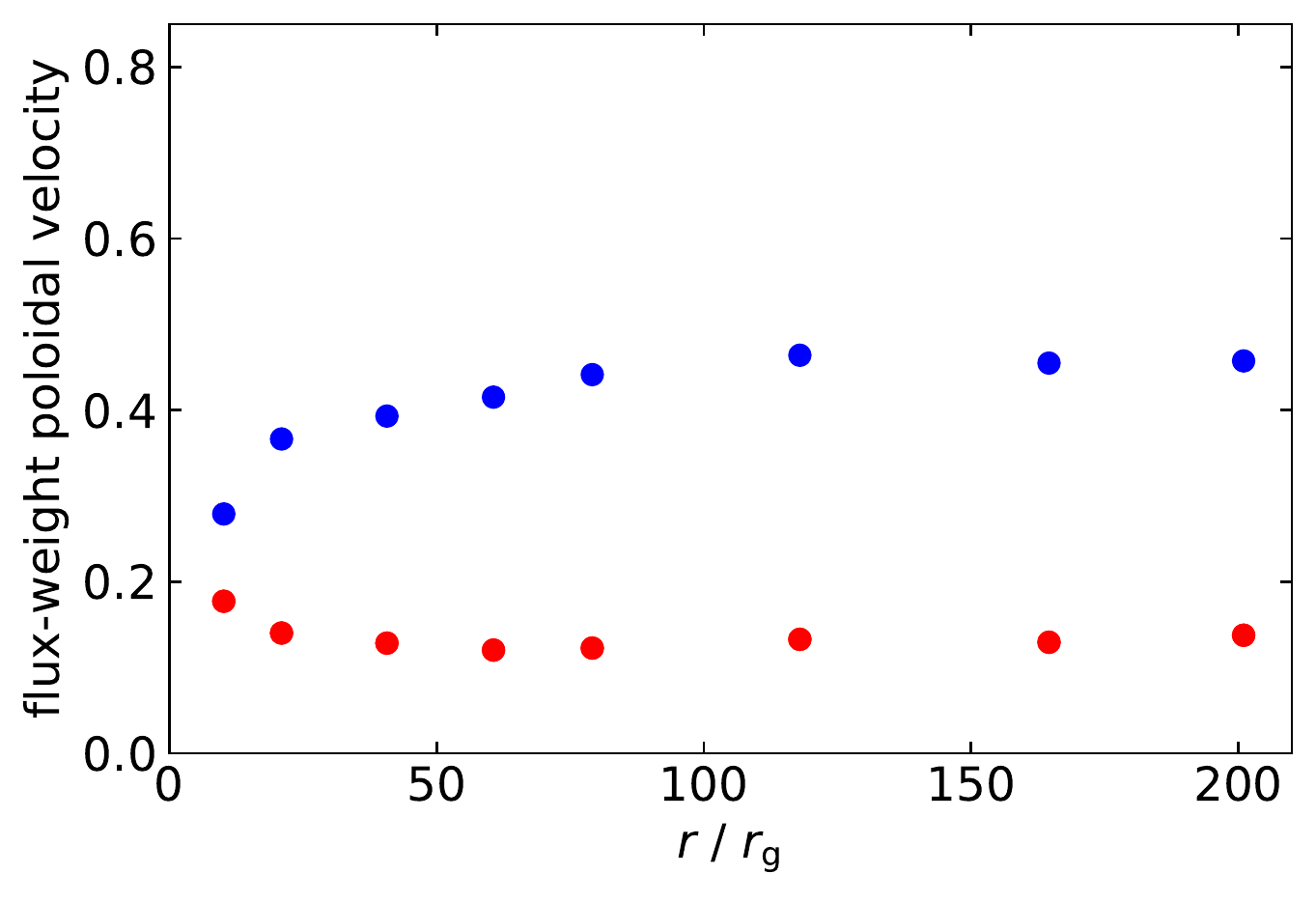}
   \caption{The mass-flux-weighted poloidal velocities of jet (blue dots) and wind (red dots). The unit of the velocity is the speed of light.} 
   \label{fig:pv}
\end{figure}

Figure~\ref{fig:pv} shows the mass-flux-weighted poloidal velocities of wind and jet. They can be roughly fitted by 
\begin{equation}
 v_{\rm p,wind}(r)\approx 0.15 c
\label{eq:windv}   
\end{equation}
and
\begin{equation}
 v_{\rm p,jet}(r)\approx 0.45 c 
\label{eq:jetv}   
\end{equation}
respectively.  
The poloidal velocity of jet in the present case is similar to that of hot accretion flow \citep{Yang2021}, where it is $\approx 0.4 c$ (SANE) or $0.5c$ (MAD). However, the poloidal velocity of wind is quite different. In the case of hot accretion flow, the wind velocity decreases with increasing radius as $\propto v_k(r)$; while in the current case it is roughly a constant. We note that the value of wind velocity we obtain is in  agreement with that obtained in the analytical work by \citet{2016MNRAS.455.1211K}, where it is found to be $\sim 0.1-0.2c$, and in the observational works by \citet{2022ApJ...931...77D} and \citet{2022A&A...668A..87V}, where it is  found to be $\sim 0.1-0.3c$. The different behavior  of the wind poloidal velocity between a super-Eddington accretion flow and a hot accretion flow is likely because that, there is  an additional acceleration by  the radiation force in the case of super-Eddington accretion. The importance of radiation acceleration is confirmed by our detailed analysis presented in section \ref{accelerationmechanism}. 

\begin{figure}
   \includegraphics[width=0.9\linewidth]{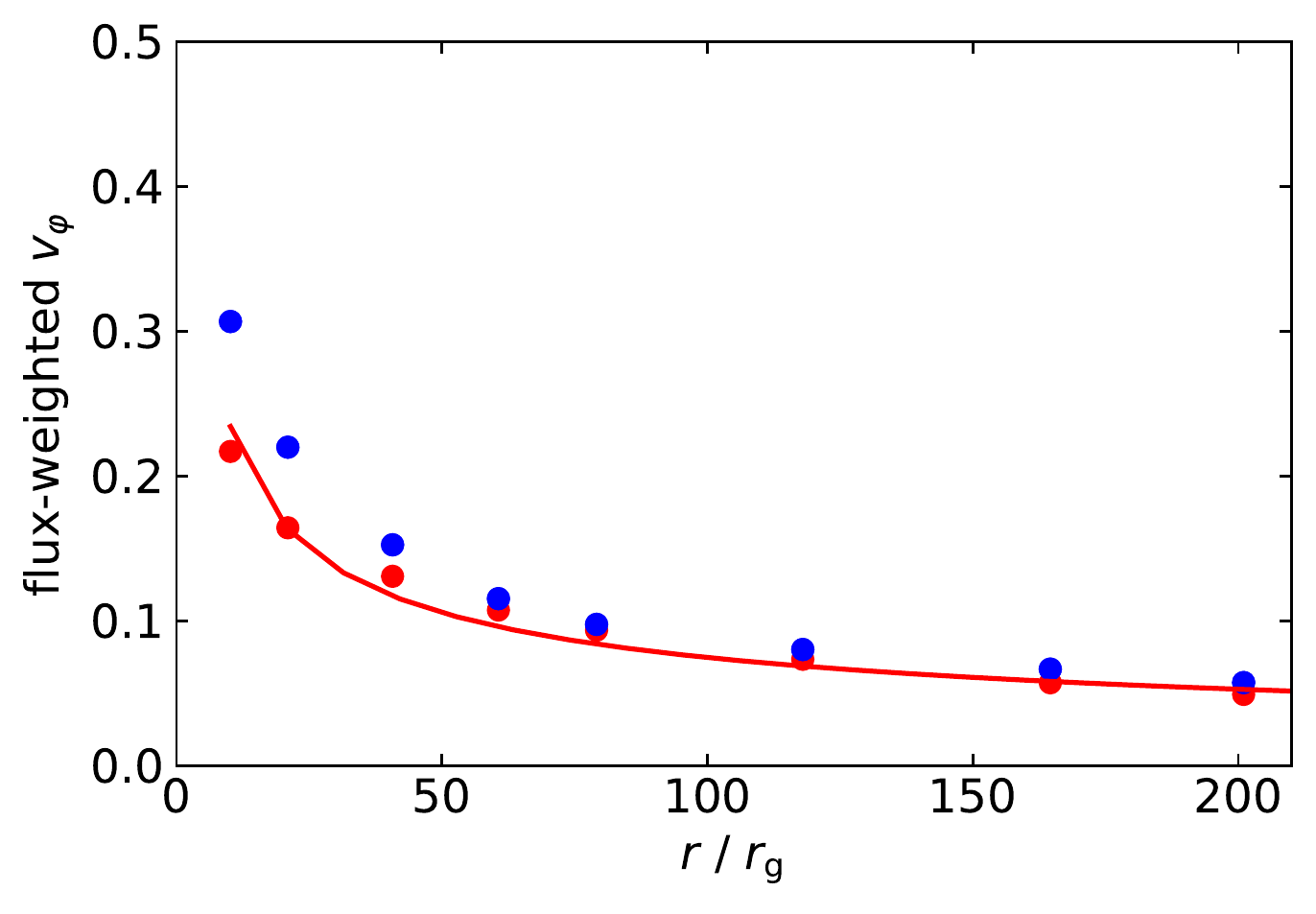}
   \caption{The mass-flux-weighted $v_{\varphi}$ of the jet (blue dots) and wind (red dots). The unit of the velocity is speed of light $c$. The red solid line shows the fitting result of {$v_{\varphi,{\rm wind}} (r)=0.75\,v_{k} (r)$}. }
   \label{fig:vphi}
\end{figure}

Figure~\ref{fig:vphi} shows the mass flux-weighted toroidal velocities of wind and jet. The toroidal velocity of both wind and jet can be fitted by 
\begin{equation}
 v_{\rm \varphi,wind(jet)}(r)\approx 0.75 v_k (r)
\label{eq:windt}   
\end{equation}
The toroidal velocity of jet is slightly larger than wind. 
By comparing Figures~\ref{fig:pv} \& \ref{fig:vphi}, we can see that the poloidal velocity of  wind is comparable to the toroidal velocity when $r \la 120\,r_{\rm g}$, but is much larger than the toroidal velocity as the radius increases. 

In the study of AGN feedback, we also need to know the power of wind. For this aim, we need to evaluate the flux-weighted  $v_{\rm p,wind}^2$ and $v_{\varphi,{\rm wind}}^2$, which are
\begin{equation}
    v_{\rm p,wind}^2= 0.025\,c^{2}
	\label{eq:vp2}
\end{equation}
\begin{equation}
    v_{\rm \varphi,wind}^2= 0.62~v_{\rm k}^2 
	\label{eq:vphi2}
\end{equation}

\begin{figure*}
   \includegraphics[width=0.31\linewidth]{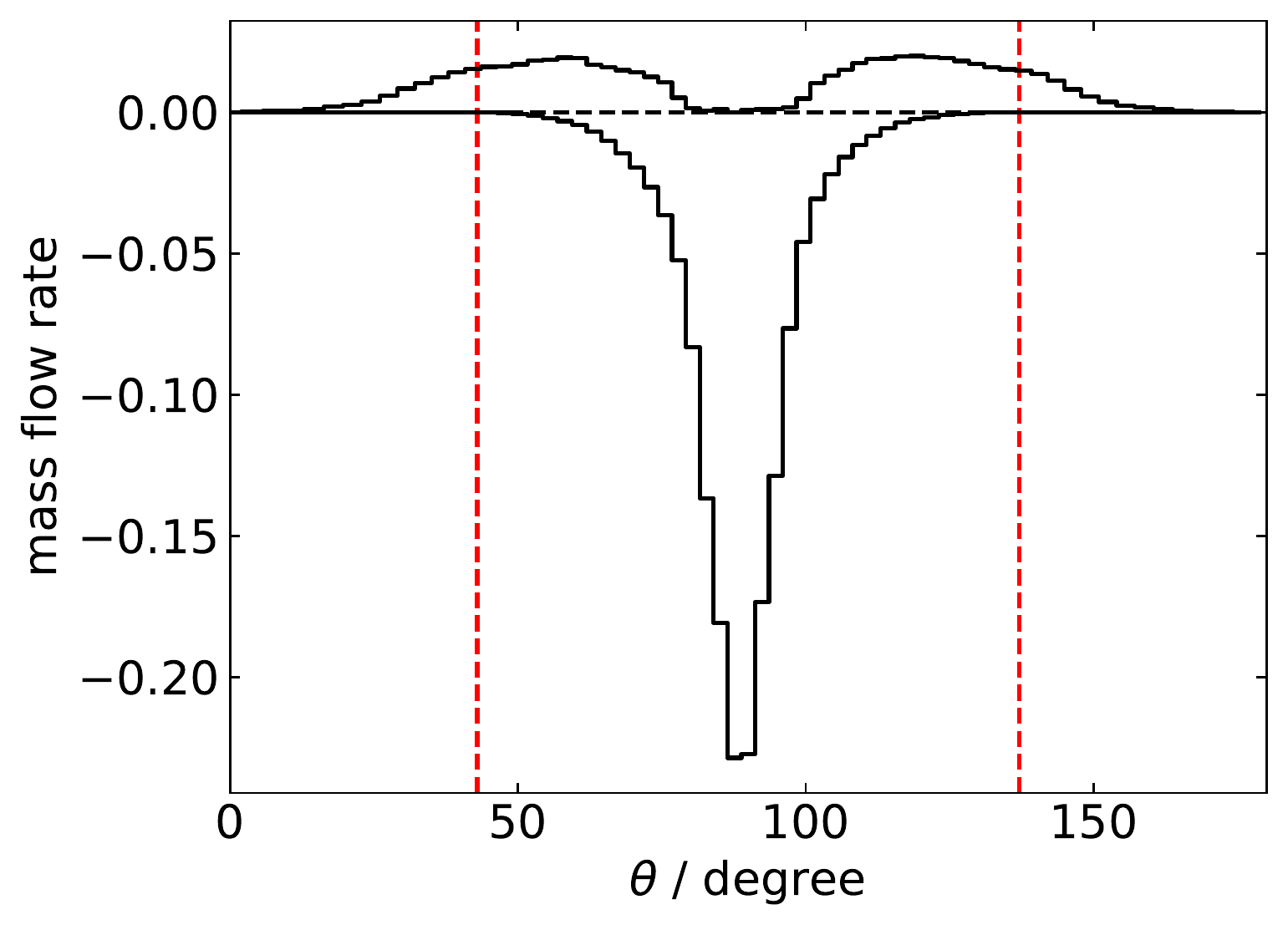}\quad\includegraphics[width=0.3\linewidth]{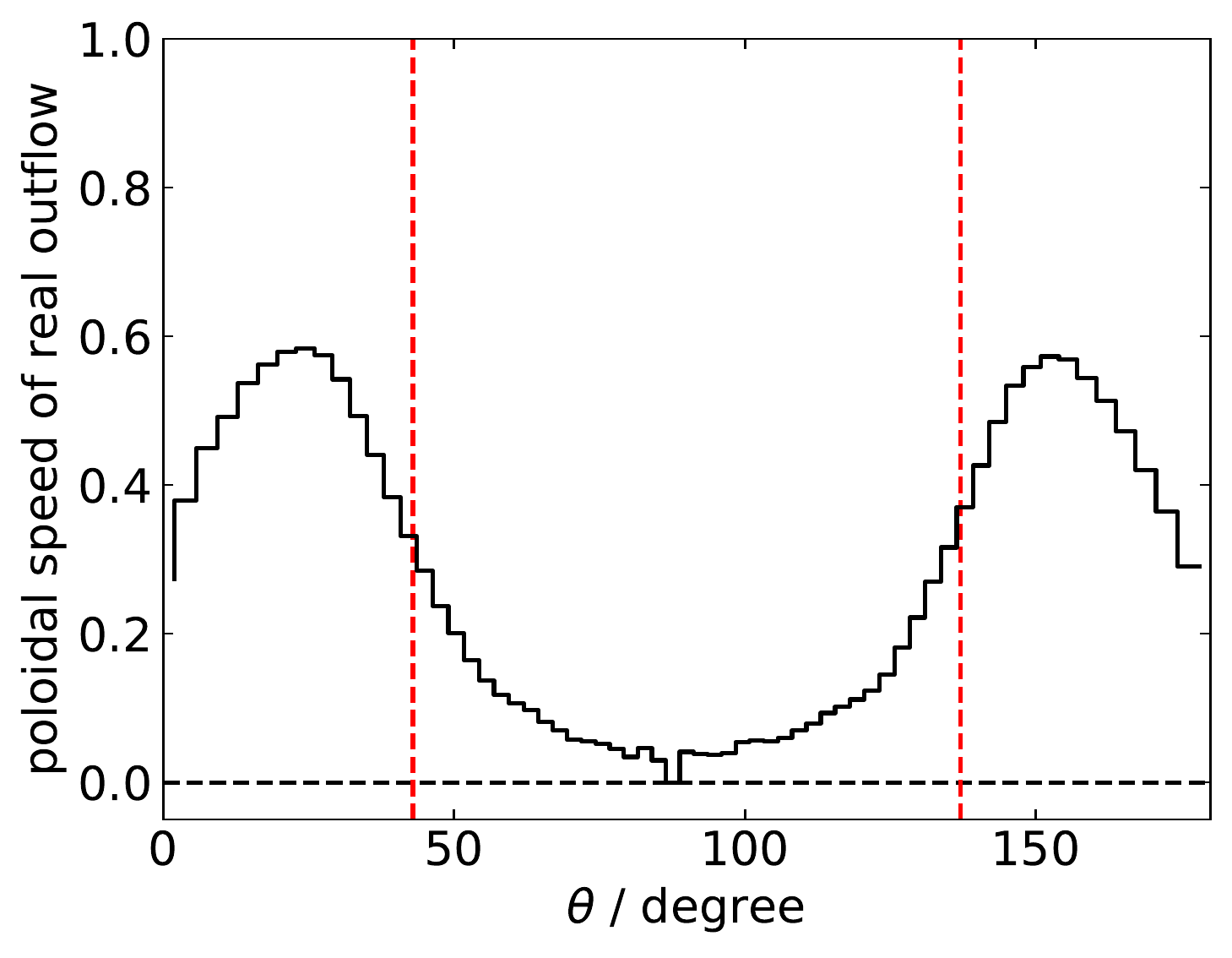}\quad\includegraphics[width=0.35\linewidth]{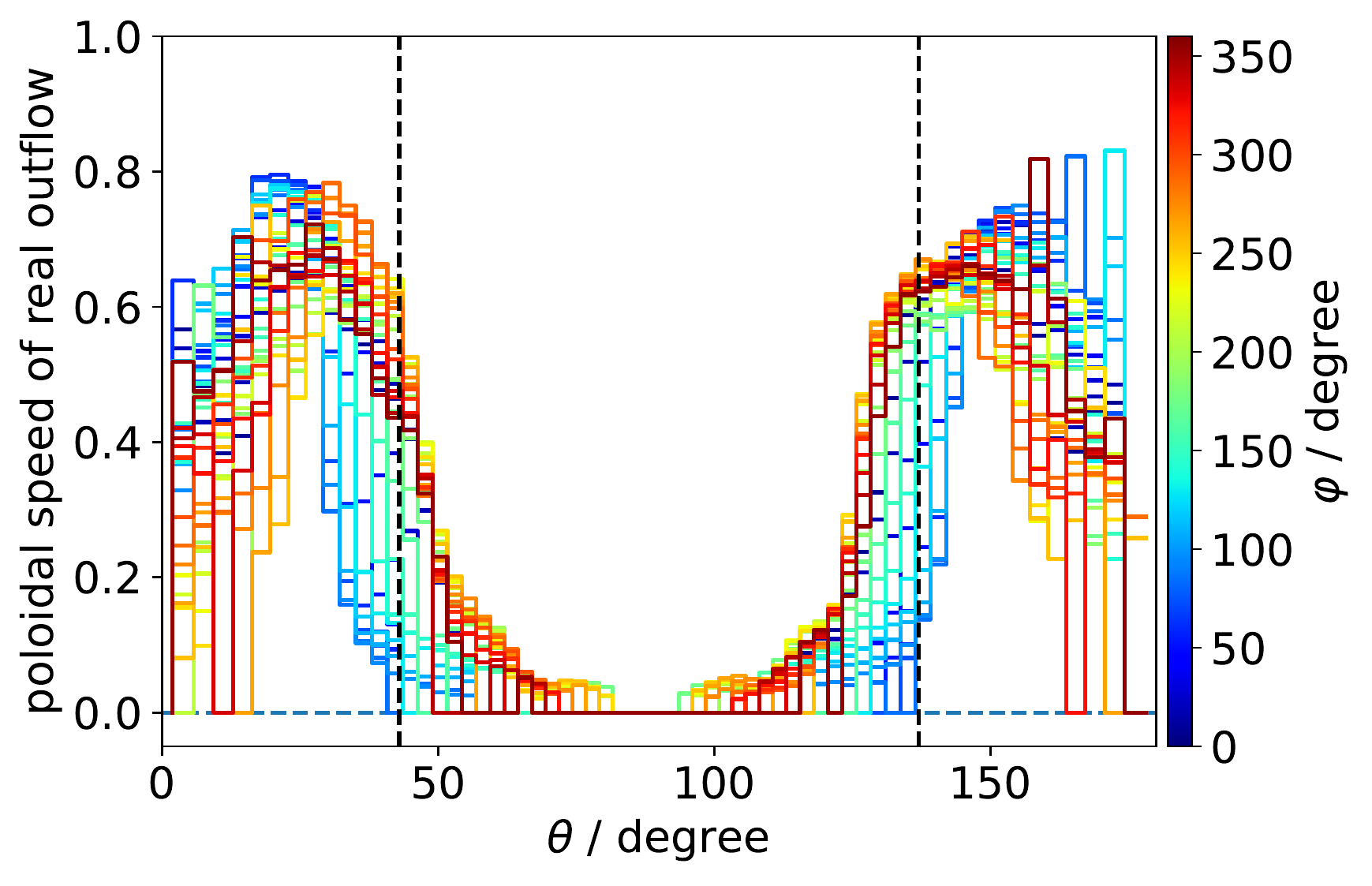}
   
   \includegraphics[width=0.31\linewidth]{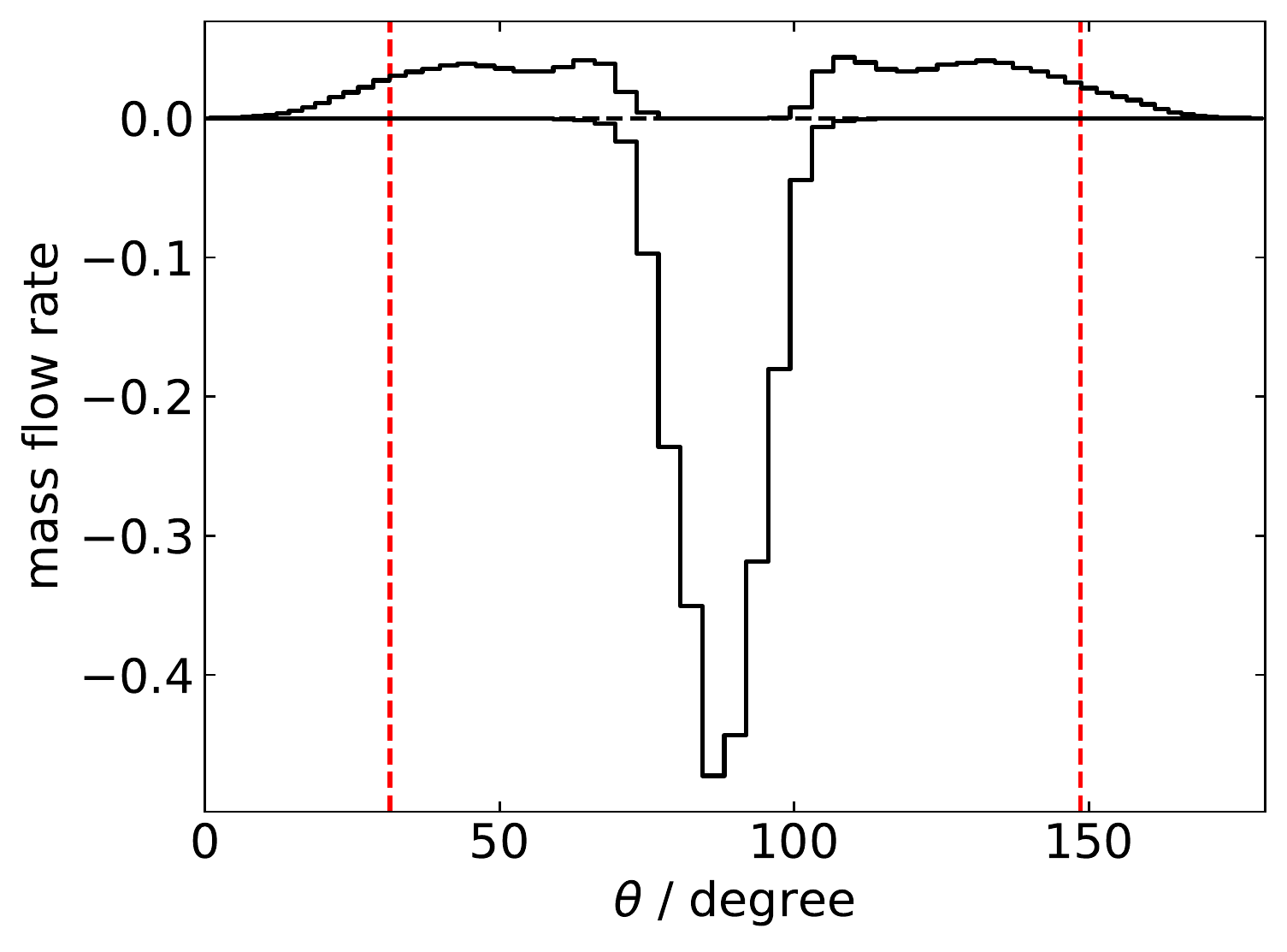}\quad\includegraphics[width=0.3\linewidth]{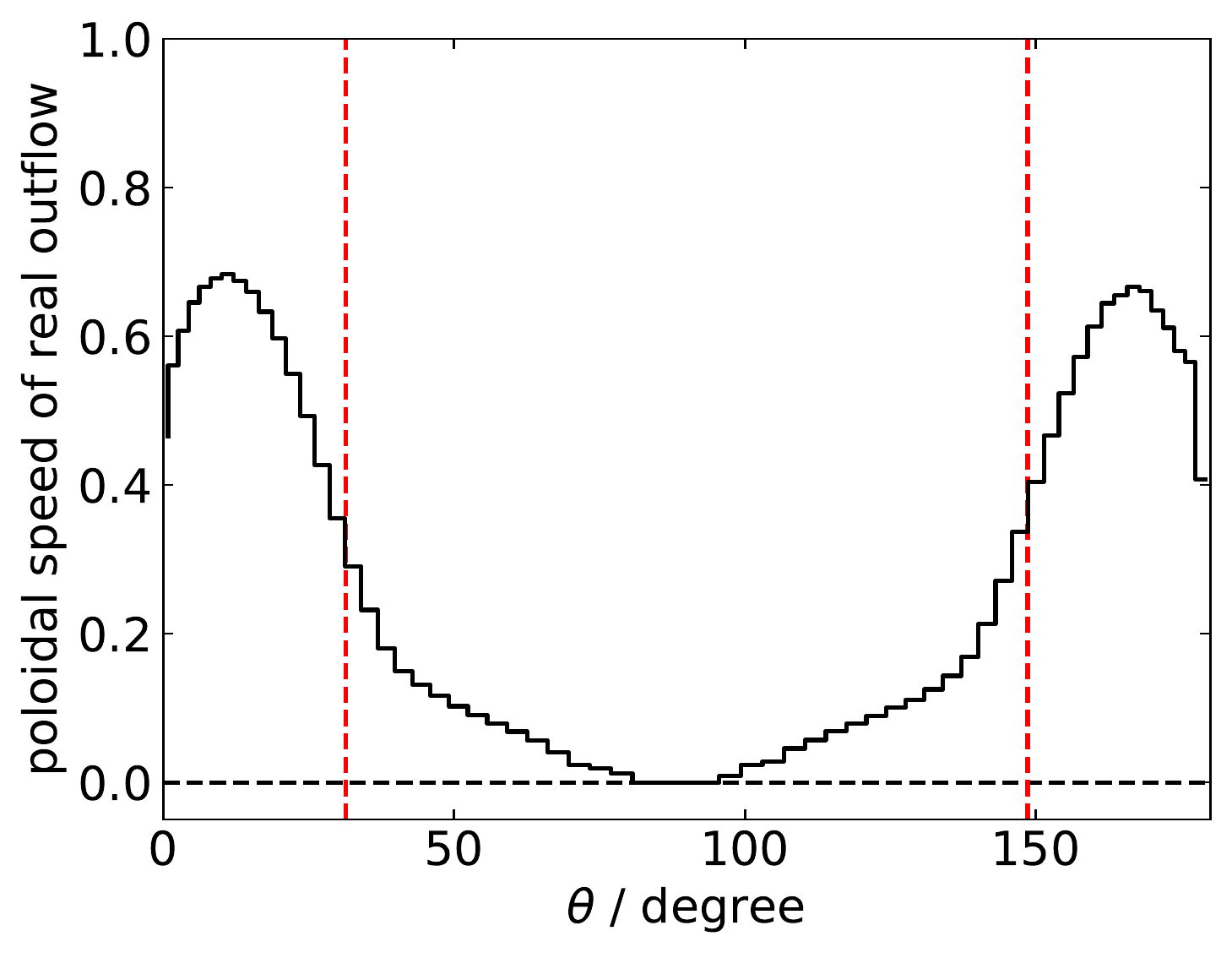}\quad\includegraphics[width=0.35\linewidth]{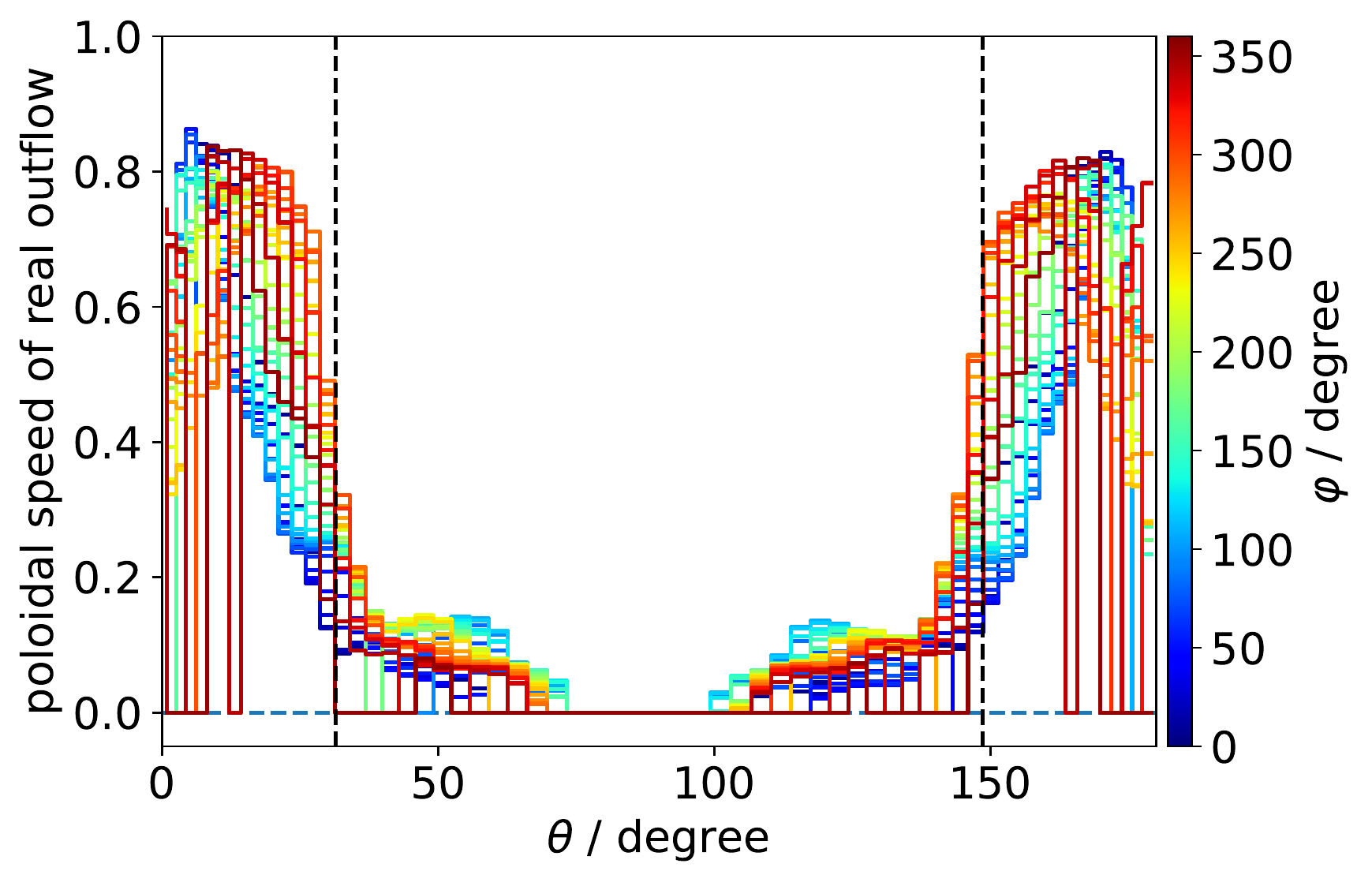}
   
   \caption{The $\theta$-dependence of the time-averaged mass fluxes of real outflow (positive) and inflow (negative) ({left}), time- and $\varphi$-averaged poloidal velocity of the real outflow ({middle}), and the time-averaged poloidal velocity at various $\varphi$ at t=8,000$\,r_{\rm {g}}/c$ ({right}). The top and bottom panels are for $r=40 \,r_{\rm {g}}/c$ and  $r=100 \,r_{\rm {g}}/c$, respectively. The vertical dashed lines denote the boundary between the wind and BZ-jet. }
   \label{fig:pvtheta}
\end{figure*}

At last, Figure~\ref{fig:pvtheta} shows the $\theta$-dependence of the mass fluxes and poloidal velocity. The inflow is mainly concentrated at equatorial plane. At $r=40\,r_{\rm g}$, the wind and inflow apparently  coexists at roughly $\theta \approx 60^{\circ} \text- 80^{\circ}$, which is not the case for $r=100\,r_{\rm g}$. 
This is because these wind particles first move almost vertically to the surface of the accretion flow before moving outward as wind. We can see that  the poloidal speed of the BZ-jet can reach to $\sim 0.88c$. 

\begin{figure*}
   \centering
   \quad\includegraphics[width=0.48\linewidth]{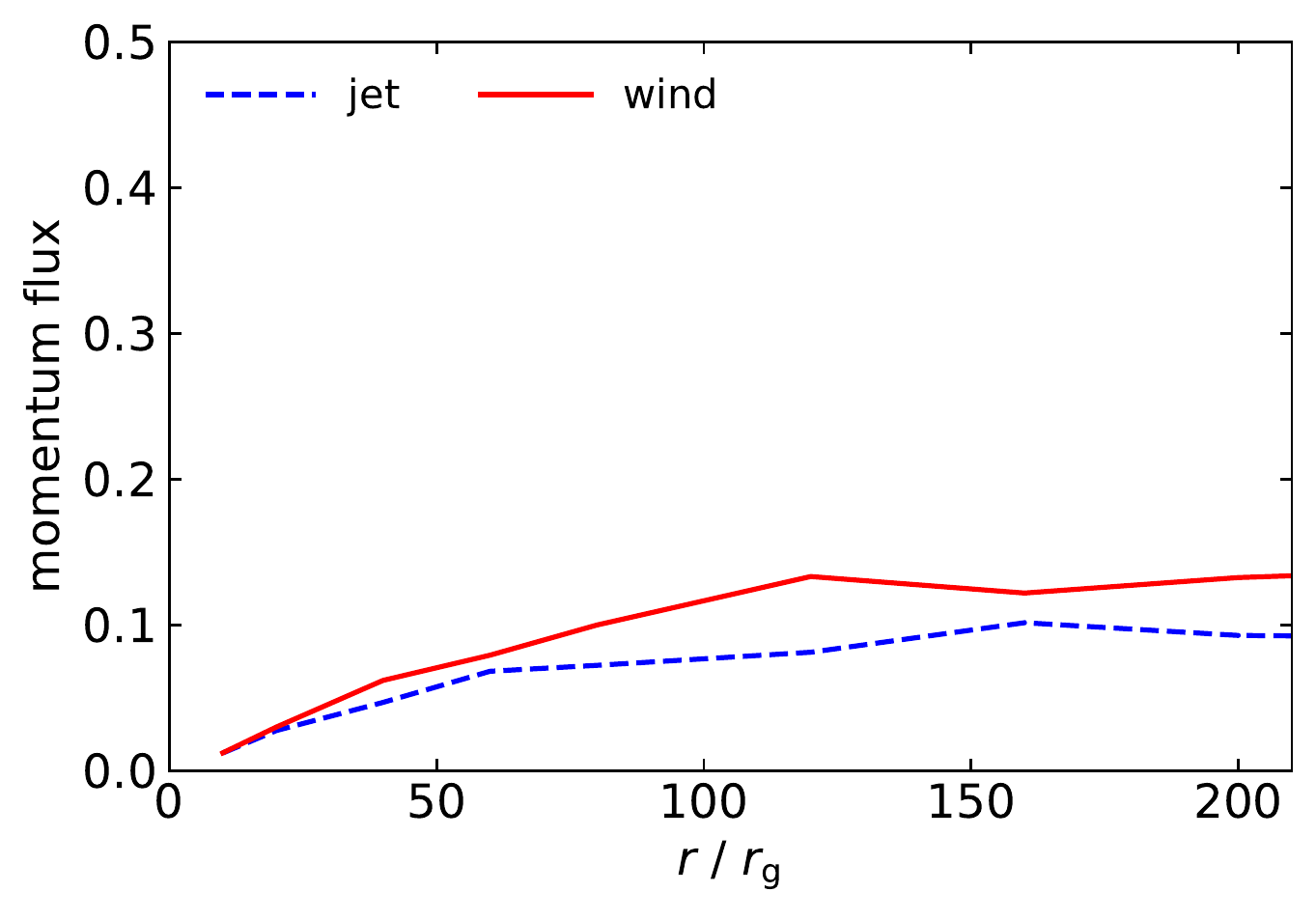}\quad\includegraphics[width=0.48\linewidth]{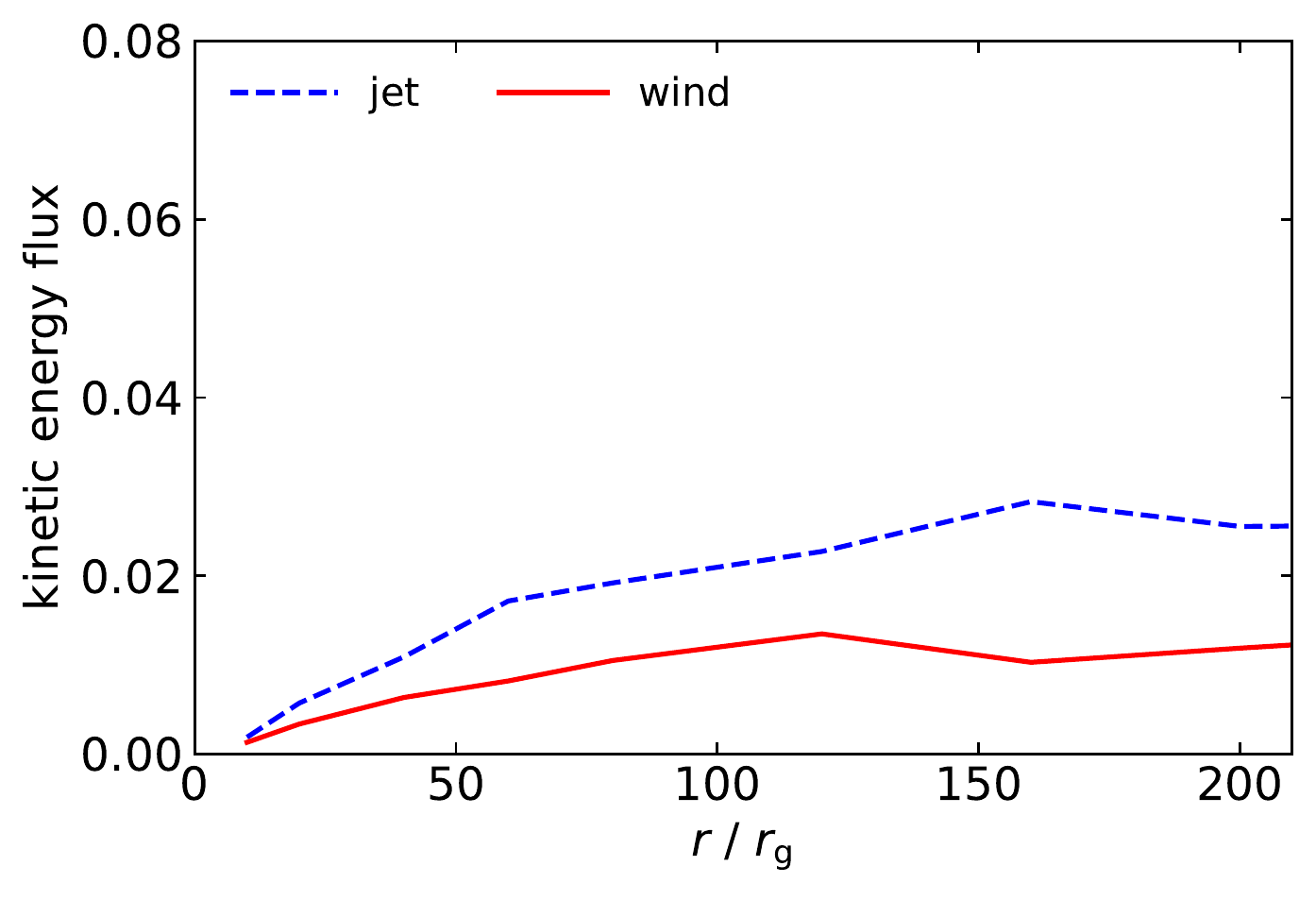}
   \caption{The radial profiles of momentum (left) and kinetic energy (right) of jet and wind. Their values have been normalized by $\dot{M}_{\rm BH}c$ and $\dot{M}_{\rm BH}c^2$ respectively. }
   \label{fig:kin}
\end{figure*}

\subsection{The energy and momentum fluxes of wind and jet}

Both jet and wind are important medium for the central AGN to affect the evolution of its host galaxy. On the one hand, they can push away the  gas surrounding the AGN via their momentum interaction; on the other hand, they can heat the surrounding gas via the energy interaction. 
The fluxes of kinetic energy and momentum of wind and jet are calculated by the following formula \citep{Yang2021}: 
\begin{equation}
\begin{split}
    \dot{E}_{\rm jet(wind)}(r)=\frac{1}{2}\int{\gamma}\rho(r, {\theta}, \varphi)v^{3}_{r}(r, \theta, \varphi)\\(r^2+a^2\cos^2\theta)\sin\theta d\theta d\varphi,
\end{split}
	\label{eq:en}
\end{equation}
\begin{equation}
\begin{split}
    \dot{P}_{\rm jet(wind)}(r)=\int{\gamma}\rho(r, {\theta}, \varphi)v^{2}_{r}(r, \theta, \varphi)\\ (r^2+a^2\cos^2\theta)\sin\theta d\theta d\varphi,
\end{split}
	\label{eq:mn}
\end{equation}
where $\gamma={1}/{\sqrt{1-v_r^2}}$. 

Figure~\ref{fig:kin} shows the radial distribution of the momentum flux (left plot) and kinetic energy flux (right plot) for jet and wind. The fluxes of momentum and kinetic energy for wind at $\sim 200\,r_{\rm g}$ are 
\begin{equation}
  \dot{P}_{\rm wind}(200r_g)\approx 0.15 \dot{M}_{\rm BH}c,
  \label{momentum200}
  \end{equation}
 \begin{equation}
  \dot{E}_{\rm wind}(200r_g)\approx 0.01 \dot{M}_{\rm BH}c^2.
  \label{kinetic200}
\end{equation}
Recall that the radiation luminosity beyond $\sim 200 r_g$ is $\sim 4.8\%\dot{M}_{\rm BH} c^2$, the kinetic energy flux of wind is $\sim 5$ times smaller than radiation while the momentum flux of wind is $\sim 3$ times larger than radiation. 

Compared with the wind launched from hot accretion flows with similar parameters (i.e., MAD around a black hole with $a=0.98$) \citep{Yang2021}, the fluxes of momentum and kinetic energy are both $\sim 2$ times smaller in unit of $\dot{M}_{\rm BH} c^2$ and $\dot{M}_{\rm BH} c$.  
The reason is likely again due to the high temperature of the hot accretion flow. 

We now compare the fluxes of energy and momentum between jet and wind, in order to evaluate their relative roles in AGN feedback. At $\sim 200r_g$, we have
\begin{equation}
  \dot{P}_{\rm wind}\approx 1.2\dot{P}_{\rm jet},
  \label{momentumratio200}
\end{equation}
 \begin{equation}
  \dot{E}_{\rm jet}\approx 2.5 \dot{E}_{\rm wind}.
  \label{kineticratio200}
\end{equation}
It is interesting to note that the above relations between jet and wind are very similar to the case of hot accretion flows  \citep{Yang2021}. 

The above results are based on simulation with a high black hole spin of $a=0.8$. How the results will change when $a$ is smaller? As we will show later, since  the power of jet in a super-Eddington accretion flow mainly comes from  extracting the spin energy of the black hole rather than from the radiation energy, if the value of $a$ were lower, the kinetic energy flux and the momentum flux of jet would be smaller. On the other hand, the fluxes of wind are expected to be not so sensitive to the black hole spin since they are launched further away from the black hole, as in the case of hot accretion flows \citep{Yang2021}. Therefore, the wind would become relatively more important than jet when $a$ is smaller. In summary, Equations (\ref{momentumratio200}-\ref{kineticratio200}) indicate that the momentum flux of wind is always larger than the jet while the energy flux of jet is at most $\sim 3$ times larger than the wind. This, combining with the fact that the solid angle of wind is orders of magnitude larger than that of jet thus wind can  deposit their energy and momentum in the surrounding medium much more efficiently than jet,  means that wind likely plays a more important role than jet in the regime of super-Eddington accretion. This result is again identical to the case of hot accretion flows \citep{Yang2021}. 

\subsection{Total energy fluxes of wind and jet}
\label{totalenergyflux}

\begin{figure}
   \includegraphics[width=0.9\linewidth]{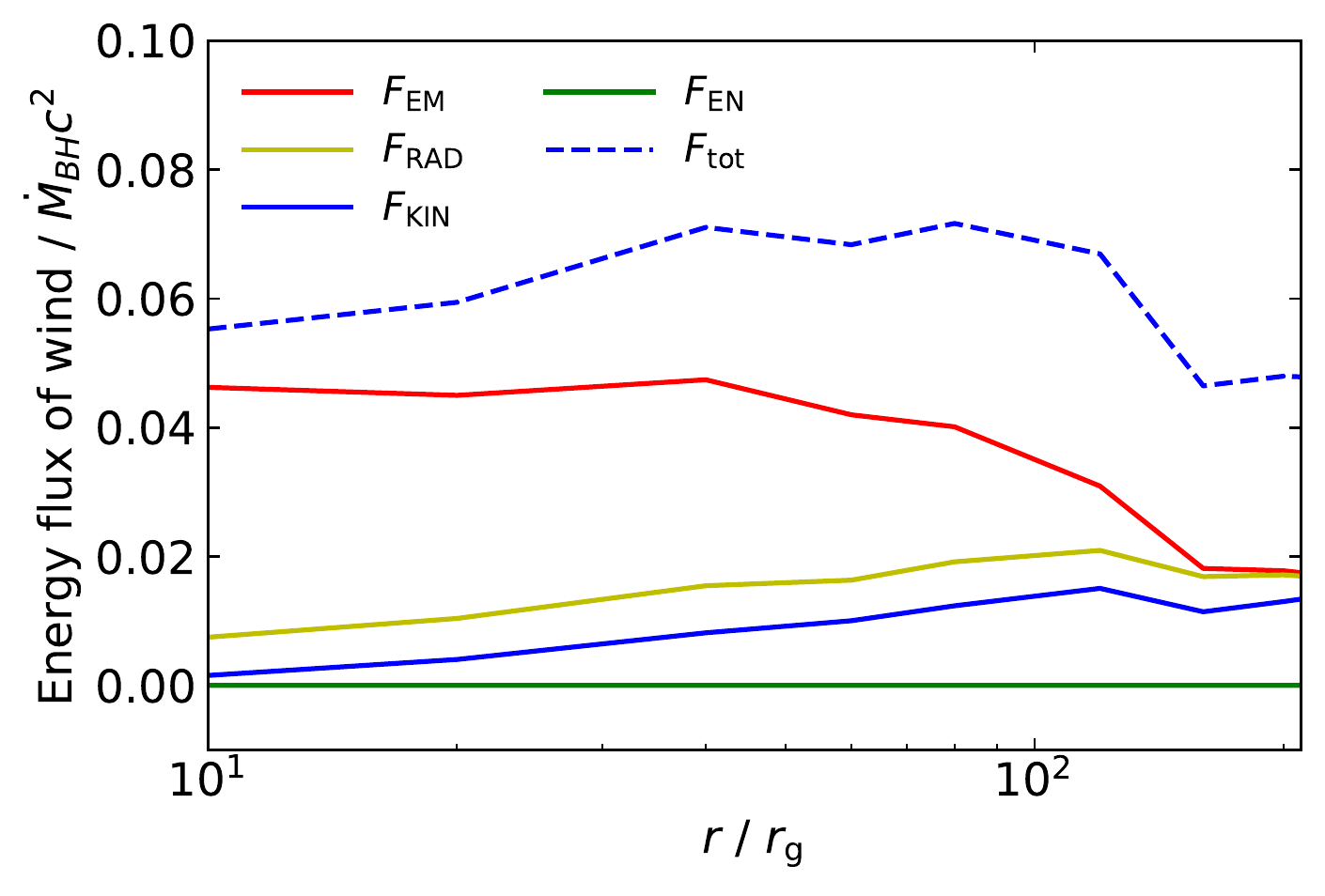}
   \caption{The radial profiles of various components of energy flux of wind normalized by $\dot{M}_{\rm BH}c^2$, including the Poynting flux $F_{\rm EM}$ (red solid), radiation flux $F_{\rm RAD}$ (yellow solid), kinetic energy  flux $F_{\rm KIN}$ (blue solid), enthalpy flux $F_{\rm EN}$ (green solid), and total energy without rest-mass energy flux $F_{\rm tot}$ (blue dashed). }
   \label{fig:windenergy}
\end{figure}

\begin{figure}
   \includegraphics[width=0.9\linewidth]{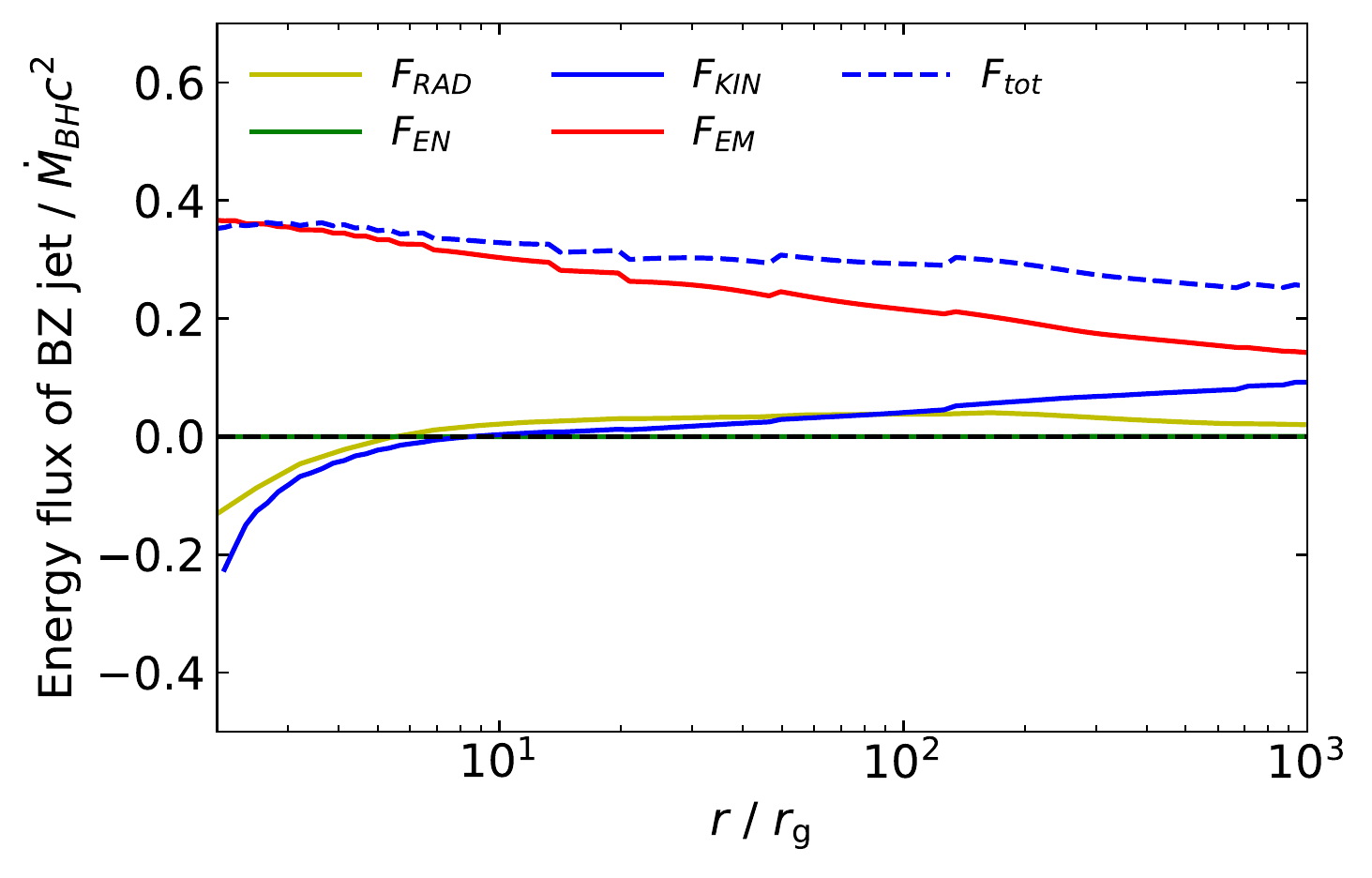}
   \caption{The radial profiles of various components of the total energy flux of the jet normalized by $\dot{M}_{\rm BH}c^2$, including the radiation flux $F_{\rm{RAD}}$ (yellow solid), the magnetic energy flux $F_{\rm{EM}}$ (red solid),  the kinetic energy flux $F_{\rm{KIN}}$ (blue solid), the  enthalpy energy flux $F_{\rm{EN}}$ (green solid), and the total energy  without the rest mass energy $F_{\rm{tot}}$ (blue dotted). }
   \label{fig:bzjet}
\end{figure}

In addition to the kinetic energy, we have also calculated all other components of the energy flux in wind and jet, including the radiation flux $F_{\rm RAD}(\equiv-\left[E_{\rm R}/3(4u^r(rad)u_{t}(rad)+\delta^{r}_{t})\right]$ (here $E_{\rm R} $ is the radiation frame radiation energy density, $u^{\mu}(rad)$ and $u_{\nu}(rad)$ are the radiation frame four-velocity), the Poynting flux $F_{\rm EM}(\equiv -\left[b^2u^ru_t-b^rb_t)\right]$, the kinetic energy flux $F_{\rm KIN} (\equiv -\left[\rho u^{r}(u_{t}+\sqrt{-g_{tt}})\right]$), the rest mass energy-subtracted energy flux  $F_{\rm tot} (\equiv -\left[(b^2+u+p+\rho)u^{r}u_{t}-b^rb_t+R^{r}_{t}\right]-{\rho}u^r$), the enthalpy flux $F_{\rm EN} (\equiv -(u+p)u^ru_t)$, and the rest-mass energy flux $F_{\rm mass}(\equiv {\rho}u^r)$. Their radial profiles are shown in Figures~\ref{fig:windenergy} \& \ref{fig:bzjet} in the cases of wind and jet, respectively. 

In the case of wind (i.e., Figure \ref{fig:windenergy}), we can see that, with the increase of $r$, the Poynting flux  $F_{\rm EM}$ decreases rapidly while the kinetic energy and radiation energy increase. 
From $r=10 r_g$ to $r=200 r_g$, the Poynting flux decreases by $\sim 0.028\, \dot{M}_{\rm BH}$, larger than the increase of kinetic energy flux. The increase of the radiation flux with radius indicates that, with increasing radius, the radiative contribution from the accretion gas is still significant. These results preliminarily suggest that the conversion of Poynting flux likely contributes to the acceleration of wind. This is confirmed by the detailed analysis of the acceleration mechanism of wind and jet to be  presented in section \ref{accelerationmechanism}.
The enthalpy flux is always very small compared with the other energy components, which is consistent with the case of SANE supercritical accretion flow around a non-spin black hole \citep{Sadowski2015b}. 

In the case of jet, we can see from Figure~\ref{fig:bzjet} that different from the wind, the Poynting flux $F_{\rm EM}$ is always the dominant component throughout the radius, while the radiation flux is roughly a constant at $r>10\,r_{\rm g}$. Same with wind, the kinetic energy flux increases with radius\footnote{The kinetic energy flux is negative at $r\la 10 r_g$, this is because the radial velocity in this region is inward.}, which is be due to the conversion of the  Poynting flux. From these analysis, we conclude that, similar to the case of wind, the jet is accelerated by the magnetic field. This conclusion is confirmed by the force analysis to be presented in Section \ref{accelerationmechanism}. 

\begin{figure}
   \includegraphics[width=0.9\linewidth]{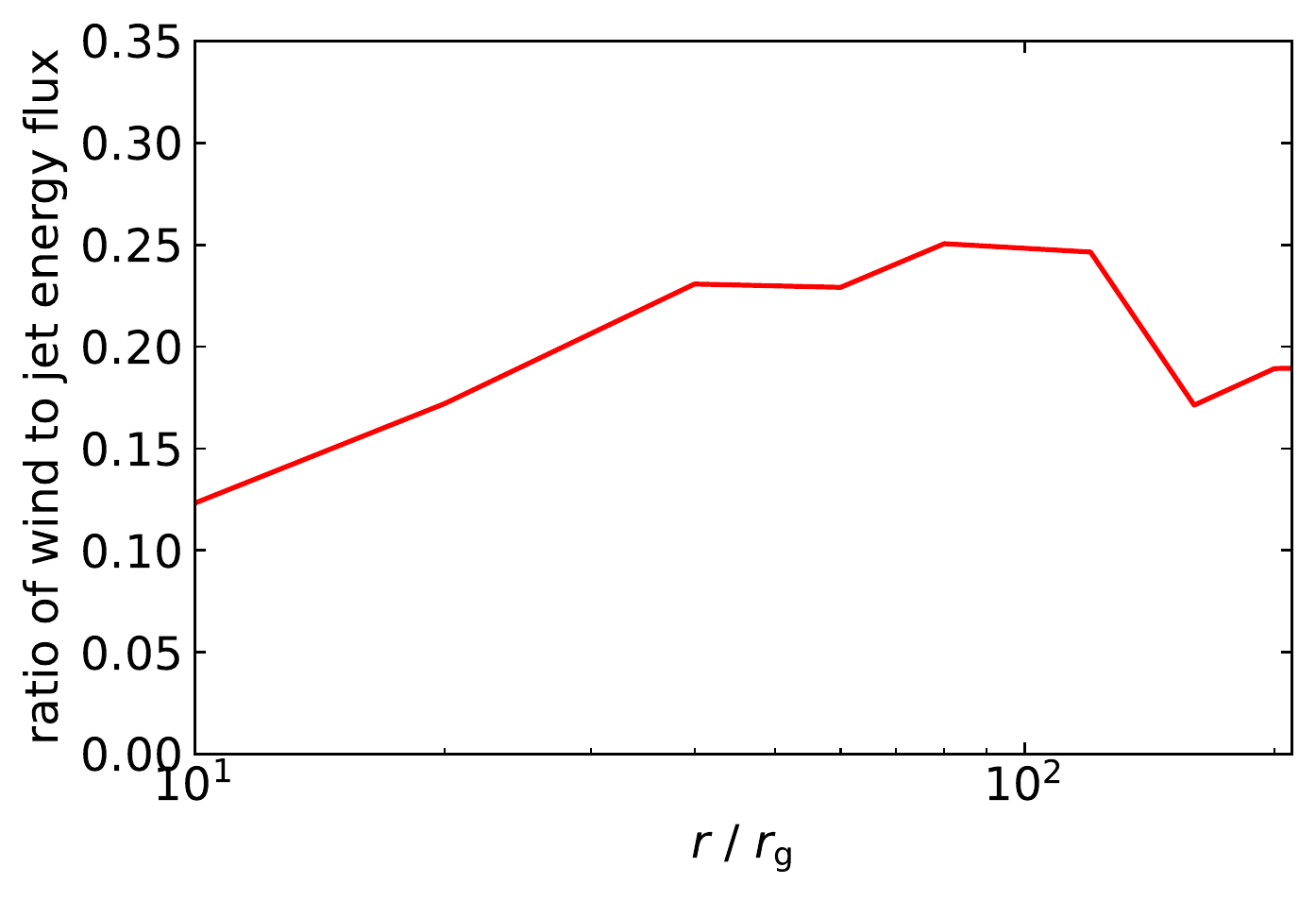}
   \caption{The radial profile of the ratio of wind to jet total energy flux.}
   \label{fig:ratio-wind2jet}
\end{figure}

To compare the total energy flux between wind and jet, we have calculated their energy flux ratio. The results is shown in Figure \ref{fig:ratio-wind2jet}. At $\sim 200\,r_{\rm g}$, the ratio of wind to jet total energy fluxes reaches
$   F_{\rm tot,wind}/F_{\rm tot,jet} \approx 19\%$. 
Interestingly, this ratio is comparable to case of hot accretion flow around a black hole with $a=0.98$, which is $10\%$ and $15\%$ for MAD and SANE respectively \citep{Yang2021}. 

\subsection{The acceleration mechanism of jet and wind}
\label{accelerationmechanism}

\begin{figure}
   \includegraphics[width=0.98\linewidth]{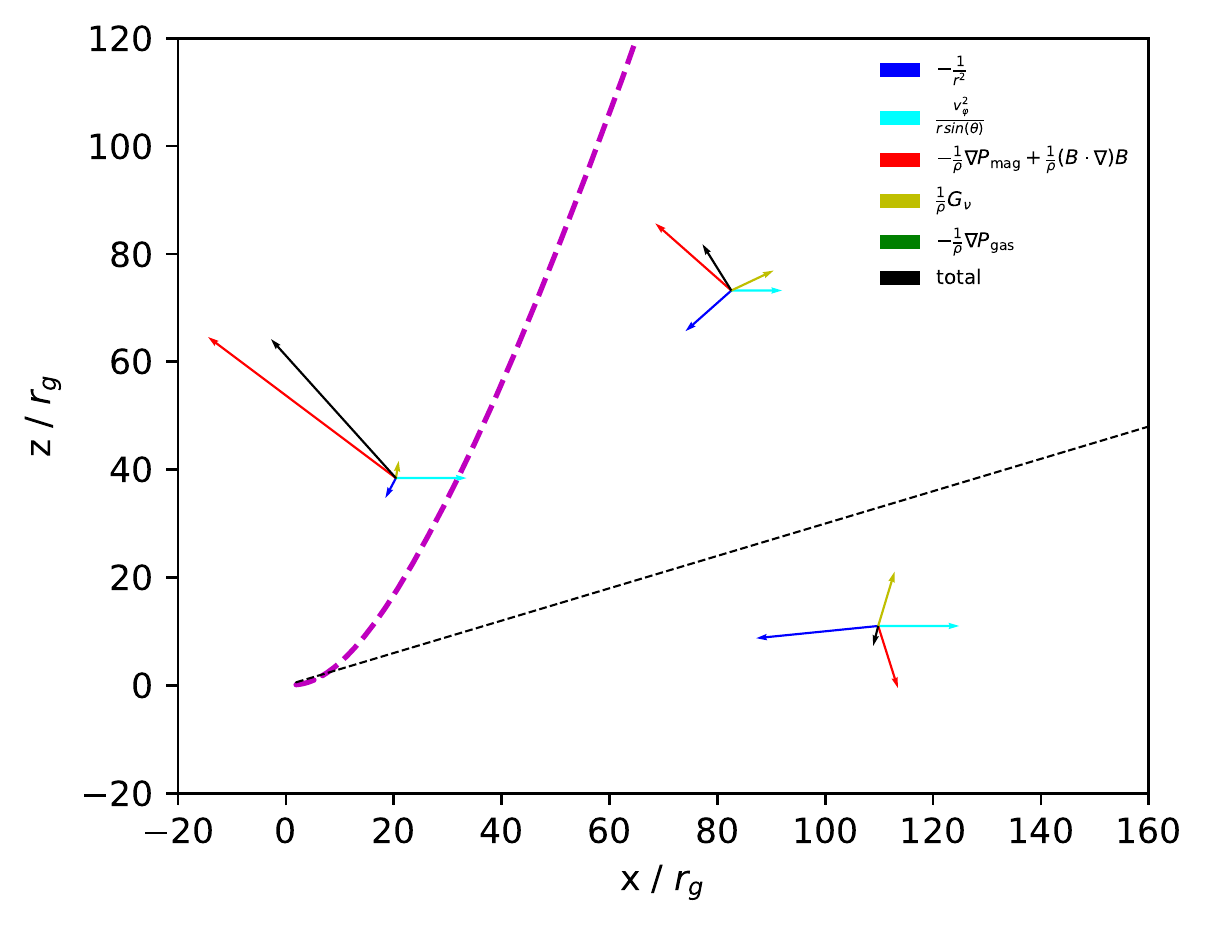}
   \caption{Force analysis at three representative locations corresponding to the BZ jet, wind, and the main body of the accretion disk. The arrows indicate force direction, whose length represents force magnitude. The purple dashed and the black dotted lines denote the boundary of the BZ-jet  and the surface of the accretion flow, respectively. }
   \label{fig:force}
\end{figure}
In the case of a hot accretion flow, it has been shown that both wind and jet are mainly accelerated by magnetic field \citep{Yuan2015}. In the present case of super-Eddington accretion, radiation becomes much more important thus potentially plays an important role in the acceleration of these outflows. A question is then among the magnetic field and radiation, which one is the dominant mechanism. In Section \ref{totalenergyflux}, by analyzing the conversion between different components of the time and spatially averaged energy flux in the jet and wind, we have obtained an initial  conclusion that these two kinds of outflow are accelerated mainly by magnetic field.  In this section, we continue to investigate this issue by adopting a more direct approach; that is, to analyze the dominant forces accelerating the jet and wind.  

Following \citet{Yuan2015}, we have chosen {\bf several} representative locations within the BZ-jet, wind, and the main body of the accretion disk to analyze the forces. The calculation is based on time-averaged data.  The results are shown in Figure \ref{fig:force}. The forces include the gravity force ($-1/r^2$), the centrifugal force ($v_{\phi}^2/r \sin{(\theta)}$), the magnetic force ($-1/ \rho \nabla P_{\rm mag}+1/ \rho (B \cdot  \nabla)B$),  the gradient of gas thermal pressure ($-1/ \rho \nabla P_{\rm gas}$), and the radiation force ($1/ \rho G_{\nu}$). The details of the calculation of the radiation force can be found in Appendix \ref{sec:force}. Since we evaluate the forces in the frame co-rotating with the flow, the
centrifugal force is also included. We find that in the jet region, the force is fully dominated by the Lorentz force. In the case of wind, the results are a bit more complicated, depending on locations. In the wind region close to the surface of the accretion flow, the optical depth is large and the radiation and gas are strongly coupled. We find that here the radiation force and the Lorentz force are comparable and both of them play the dominant roles in the acceleration of wind. Well above the main body of the accretion flow, the optical depth becomes smaller. In this location, compared to the Lorentz force, the radiation force becomes less important in accelerating the wind and is the secondarily important force. The direction of the Lorentz force in the wind and jet region is significantly different from that in the case of hot accretion flow shown in Fig. 13 of \citet{Yuan2015}. We find that the main reason for such a discrepancy is that the configuration of the magnetic field between the current super-Eddington MAD accretion and the SANE case there is different. This force on one hand can accelerate the wind and jet, on the other hand, it can also collimate them. 

\section{Summary}

In this paper, we investigate the properties of wind and jet launched from a magnetically arrested super-Eddington accretion flow around a black hole with spin $a=0.8$ based on radiation magnetohydrodynamic simulations. Following our previous works of studying wind and jet in hot accretion flows \citep{Yuan2015,Yang2021}, we adopt the ``virtual particle trajectory'' method which can loyally reflect the motion of the fluid elements thus judge whether they are turbulence (``false'' outflow) or real outflow (i.e., wind and jet). We define the boundary between jet and wind as the magnetic field line whose foot point is rooted at the black hole ergosphere with $\theta=90^{\circ}$. Under this definition, our jet is exactly the Blandford-Znajeck jet, which is suggested by the excellent agreement of the jet morphology between theoretical prediction and observations \citep{Yang2022}. Some other different definitions of jet and wind and the calculation method of their properties exist in literature. In some papers, for example,  wind and jet are defined as outflowing gas with Bernoulli parameter $Be>0$. The problem with this definition is that $Be$ is usually not a constant along the propagating outflow thus not a good indicator of real outflow. When calculating the properties of outflow such as the mass flux, people often do time-average to the simulation data first in order to filter out the turbulence. But since wind is instantaneous, in this way some real outflow will be ``cancelled'' thus underestimate the flux.  The main results of the present paper are summarized as follows.

\begin{itemize}
    \item The radial profiles of mass fluxes of wind and jet are obtained and presented in Figure~\ref{fig:wind} and Equations~ (\ref{eq:wind}) \& (\ref{jetflux}). The mass flux of the wind increases with increasing radius. The maximum value of radius in Equation (\ref{eq:wind}) should be equal to the outer boundary of the accretion flow, which is Bondi radius in the case of AGN.  The mass flux of the jet however roughly reaches a saturated value at $\sim 60 r_g$. These results are similar to the case of wind and jet launched from hot accretion flows. 
    
    \item For comparison, we have also calculated the mass flux of wind and jet following the ``time-average'' method often adopted in literature. We find that the mass flux of wind is $2-6$ times smaller than our result, depending on radius; while the mass flux of jet is a factor $\sim 2$ larger than our result. The discrepancy for the latter is due to the difference of the definition of jet.   
    
    \item In addition to mass flux, another most important quantity for the AGN feedback study is the velocity of outflow. The radial profiles of  mass-flux-weighted poloidal speed of wind and jet  are presented in Figures~\ref{fig:pv} and Equations (\ref{eq:windv}) \& (\ref{eq:jetv}) respectively. Compared to the case of hot accretion flow, the jet velocity is similar \citep{Yuan2015,Yang2021}; however,  the velocity of wind is higher and does not change with increasing radius. This is because of the additional acceleration by radiation force in the case of super-Eddington accretion. The radial profile of the toroidal speed of wind and jet are presented in Figure \ref{fig:vphi} and  Equation~\ref{eq:windt}. 
    
\item The impact of the outflow to the ISM in the host galaxy is via the momentum and energy interaction. The fluxes of momentum and kinetic energy of jet and wind are presented in  Figure~\ref{fig:kin} and the comparisons between jet and wind 
are presented in Equations (\ref{momentumratio200})-(\ref{kineticratio200}). 
The momentum flux of wind is larger than jet, while the total energy flux of jet is about 3 times larger than that of wind. Given that: 1) we are considering a MAD around a black hole with a large spin, 2) both magnetic field and spin are in favor of jet power, and 3) the opening angle of wind is much larger than jet,  we conclude that the wind likely plays a more important role than jet in AGN feedback. Such a conclusion is identical to the case of wind and jet launched from a hot accretion flow \citep{Yuan2015,Yang2021}.
    
\item The radial profiles of  different components of the energy fluxes of jet and wind are shown in Figures \ref{fig:windenergy} \& \ref{fig:bzjet}. The power of jet and wind are dominated by Poynting flux close to the black hole. The Poynting flux is gradually converted into kinetic energy when the jet and wind propagate outward. 
In the case of jet, the Poynting flux is always dominant. 

\item To investigate the acceleration mechanism of wind and jet, we have analyzed the forces acting in the jet and wind region and the results are presented in Figure \ref{fig:force}. We find that the Lorentz force is the dominant force in both the wind and jet cases so magnetic field rather than radiation is the dominant mechanism, as the disk has reached the MAD state. This conclusion is same with the case of hot accretion flow. The radiation force and centrifugal force play secondarily important roles.
    
\end{itemize}

\section*{Acknowledgements}
We thank Dr. Defu Bu for the useful discussions and comments on the paper. HY and FY are supported in part by the Natural Science Foundation of China (grants 12133008, 12192220, and 12192223). LD and TK are supported by the National Natural Science Foundation
of China (grant HKU12122309) and the Hong Kong Research Grants Council (grants HKU27305119, HKU17305920, HKU 17314822). The calculations have  made use of the Tianhe-II supercomputer and the High Performance Computing
Resources in the Core Facility for Advanced Research Computing
at Shanghai Astronomical Observatory and that by the ITS at HKU. 

\section*{Data Availability}
The simulation data underlying this article will be shared on reasonable request to the corresponding author.



\bibliographystyle{mnras}
\bibliography{yang} 




\appendix
\section{The calculation of the force}\label{sec:force}
As in \citet{Moller2015}, we have radiative four-force
\begin{equation}
\begin{split}
    G^{\mu}=&\rho (\kappa_{\rm a} +\kappa_{\rm s})R^{\mu \nu} u_{\nu}-\rho(\kappa_{\rm s} R^{\alpha \beta}u_{\alpha}u_{\beta}+ \kappa_{\rm a}4\pi B )u^{\mu}\\
    &+G^{\mu}_{\rm comp},
\end{split}
	\label{eq:radforce0}
\end{equation}
where $\kappa_{\rm a}$ and $\kappa_{\rm s}$ are the opacities for the free-free absorption and Thomson-scattering,

\begin{equation}
   \kappa_{\rm a}=6.4\times10^{22}\rho T_{\rm gas}^{-7/2}\, \rm{cm}^{2}/g ,
	\label{eq:kabs}
\end{equation}
\begin{equation}
 \kappa_{\rm s}=0.34\, \rm{cm}^{2}/g,
	\label{eq:ksca}
\end{equation}
where $T_{\rm gas}$ is the gas temperature, and $B=a_{\rm rad}T_{\rm gas}^{4}/4\pi$, $a_{\rm rad}$ is the radiation constant.
The force of thermal Comptonization is,
\begin{equation}
\begin{split}
    G^{\mu}_{\rm comp}=&-\kappa_{\rm s}\rho \widehat{E} [\frac{4k(T_{\rm gas}-T_{\rm rad})}{m_e c^2}]\times[1+3.683\frac{k T_{\rm gas}}{m_e c^2}+4\frac{k T_{\rm gas}}{m_e c^2}^{2}]\\
    &\times[\frac{k T_{\rm gas}}{m_e c^2}]^{-1} u^{\mu},
\end{split}
	\label{eq:radforce1}
\end{equation}
where $\widehat{E}=R^{\mu\nu}u_{\mu}u_{\nu}$ is the comoving frame radiation energy density, $T_{\rm rad}=(\widehat{E}/a_{\rm rad})^{1/4}$ is the radiative temperature. Then we would have the radiation force $G_{\nu}$ in the Locally Non-Rotating Frames (LNRF).









\bsp	
\label{lastpage}
\end{document}